\documentclass[a4paper]{llncs}
\usepackage{amsmath}
\usepackage{amssymb}
\usepackage{pifont}
\usepackage[table]{xcolor}
\usepackage{rotating}
\usepackage{subfigure}
\usepackage{booktabs}
\usepackage{multirow}
\usepackage[linesnumbered,vlined,ruled,commentsnumbered]{algorithm2e}

\usepackage{tikz}
\usetikzlibrary{arrows,shapes,automata,matrix,positioning,chains,
calc,shapes.arrows,patterns}
\tikzset{
double -latex/.style args={#1 colored by #2 and #3}{    
    -latex,line width=#1,#2,
    postaction={draw,-latex,#3,line width=(#1)/3,shorten <=(#1)/4,shorten >=4.5*(#1)/3},
  },
  double arrow/.style args={#1 colored by #2 and #3}{
    -stealth,line width=#1,#2, % first arrow
    postaction={draw,-stealth,#3,line width=(#1)/3,
                shorten <=(#1)/3,shorten >=2*(#1)/3}, % second arrow
  }
}

\newcommand{\SMAUG}{\textsc{SMAUG}}
\newcommand{\gesture}[1]{\texttt{#1}}
\newcommand{\ctableyesS}{\cellcolor{green!15}}
\newcommand{\ctablenoS}{\cellcolor{red!15}}
\newcommand{\ctablemaybeS}{\cellcolor{yellow!15}}
\newcommand{\ctableyes}{\ctableyesS yes}
\newcommand{\ctableno}{\ctablenoS no}
\newcommand{\ctablemaybe}{\ctablemaybeS some}
\newcommand{\N}{\mathbb{N}}
\newcommand{\R}{\mathbb{R}}
\newcommand{\DTW}{\operatorname{DTW}}
\newcommand{\touchdataS}{\mathcal{T}}
\newcommand{\gyrodataS}{\mathcal{G}}
\newcommand{\laccdataS}{\mathcal{A}}
\newcommand{\motiondataS}{\mathcal{S}}
\newcommand{\fusiondataS}{\mathcal{F}}
\newcommand{\metadata}{\mathcal{M}}
\newcommand{\touchdata}[1]{\touchdataS^{(#1)}}
\newcommand{\gyrodata}[1]{\gyrodataS^{(#1)}}
\newcommand{\laccdata}[1]{\laccdataS^{(#1)}}
\newcommand{\motiondata}[1]{\motiondataS^{(#1)}}
\newcommand{\fusiondata}[1]{\fusiondataS^{(#1)}}
\newcommand{\round}{r}
\newcommand{\user}{\mathsf{U}}
\newcommand{\unknownuser}{\user^*}
\newcommand{\cmark}{\ding{51}}
\newcommand{\xmark}{\ding{55}}
\newcommand{\systemparameter}{\mathsf{P}}
\newcommand{\sprounds}{{\ensuremath{\systemparameter_E}}}
\newcommand{\sprepeat}{\ensuremath{\systemparameter_V}}
\newcommand{\spfaultweightmult}{\ensuremath{\systemparameter_{W^*}}}
\newcommand{\spfaultweightadd}{\ensuremath{\systemparameter_{W^+}}}
\newcommand{\spfaultnumbermult}{\ensuremath{\systemparameter_{F^*}}}
\newcommand{\spfaultnumberadd}{\ensuremath{\systemparameter_{F^+}}}
\newcommand{\tierset}{T}
\newcommand{\tierone}{{\tierset 1}}
\newcommand{\tiertwo}{{\tierset 2}}
\newcommand{\tierthree}{{\tierset 3}}

\newcommand{\sptierone}{\ensuremath{\systemparameter_\tierone}}
\newcommand{\sptiertwo}{\ensuremath{\systemparameter_\tiertwo}}
\newcommand{\sptierthree}{\ensuremath{\systemparameter_\tierthree}}
\newcommand{\spmotionstartoffset}{\ensuremath{\systemparameter_{OS}}}
\newcommand{\spmotionendoffset}{\ensuremath{\systemparameter_{OE}}}
\newcommand{\weightsset}{\mathcal{W}}
\newcommand{\threshold}{\Theta}
\newcommand{\algTemplateGeneration}{\operatorname{TemplateGeneration}}

\newcommand{\algFeatureExtraction}{\operatorname{FeatureExtraction}}

\newcommand{\metadataset}{\mathcal{M}}
\newcommand{\touchdataset}{\mathcal{T}}
\newcommand{\motiondataset}{\mathcal{S}}
\newcommand{\algDataGathering}{\operatorname{DataGathering}}
\newcommand{\algPostProcessing}{\operatorname{PostProcessing}}
\newcommand{\algComputeWeights}{\operatorname{ComputeWeights}}
\newcommand{\algThreshold}{\operatorname{Thresholds}}
\newcommand{\testmetadata}{\widetilde{M}}
\newcommand{\testtouchdata}[1]{\widetilde{\touchdataS}^{(#1)}}
\newcommand{\testmotiondata}[1]{\widetilde{\motiondataS}^{(#1)}}
\newcommand{\testlaccdata}[1]{\widetilde{\laccdataS}^{(#1)}}
\newcommand{\testgyrodata}[1]{\widetilde{\gyrodataS}^{(#1)}}
\newcommand{\True}{\textsf{true}}
\newcommand{\False}{\textsf{false}}
\newcommand{\faultindicator}{\mathcal{I}_F}
\newcommand{\weightindicator}{\mathcal{I}_W}
\newcommand{\faultscount}{\widetilde{\faultindicator}}
\newcommand{\faultsweight}{\widetilde{\weightindicator}}
\newcommand{\decision}{\operatorname{dec}}
\newcommand{\authtry}{t}
\newcommand{\algTestFeatureExtraction}{\operatorname{VerificationFeatureExtraction}}

\newcommand{\algValidation}{\operatorname{Validation}}
\newcommand{\algVerification}{\operatorname{Verification}}
\newcommand{\systemtimenano}{\nu}
\newcommand{\pressure}{P}
\newcommand{\size}{S}
\newcommand{\TouchDown}{\texttt{TOUCH\_DOWN}}
\newcommand{\TouchPDown}{\texttt{TOUCH\_POINTER\_DOWN}}

\newcommand{\TouchPUp}{\texttt{TOUCH\_POINTER\_UP}}
\newcommand{\TouchMove}{\texttt{TOUCH\_MOVE}}

\newcommand{\velocityX}{V^{(X)}}
\newcommand{\velocityY}{V^{(Y)}}
\newcommand{\accelerationX}{A^{(X)}}
\newcommand{\accelerationY}{A^{(Y)}}
\newcommand{\curvature}{C}
\newcommand{\direction}{D}
\newcommand{\feature}{\phi}
\newcommand{\featureset}{\mathbb{F}}
\newcommand{\stroke}{s}
\newcommand{\AM}{\operatorname{AM}}
\newcommand{\RMS}{\operatorname{RMS}}
\newcommand{\VAR}{\operatorname{VAR}}
\newcommand{\StDev}{\operatorname{StDev}}
\newcommand{\MAD}{\operatorname{MAD}}
\newcommand{\Skew}{\operatorname{Skew}}
\newcommand{\Kurt}{\operatorname{Kurt}}

\newcommand{\PCC}{\operatorname{PCC}}
\newcommand{\sensorval}{\operatorname{sensorval}}
\newcommand{\dtw}{d}
\newcommand{\bestround}{{\round^\star}}
\newcommand{\Median}{\operatorname{Median}}
\newcommand{\anygesture}{\gesture{G}}
\newcommand{\anyunknowngesture}{\anygesture^\gesture{*}}
\newcommand{\faultcontainer}{\mathcal{E}}
\newcommand{\comparisontypes}{\mathfrak{C}}
\newcommand{\LB}{\texttt{LB}}
\newcommand{\UB}{\texttt{UB}}
\newcommand{\EQ}{\texttt{EQ}}
\newcommand{\comparison}{c}
\newcommand{\compfeature}{\feature^*}
\newcommand{\compfeatureset}{\featureset^*}
\newcommand{\compfeaturesetg}{\compfeatureset_\anygesture}
\newcommand{\compfeaturesets}{\compfeatureset_\stroke}
\newcommand{\strokes}{{n_\stroke}}
\newcommand{\faultcontainerg}{\faultcontainer_0}
\newcommand{\faultcontainers}{\faultcontainer_\stroke}
\newcommand{\weight}{\omega}
\newcommand{\testfaultindicator}{\widetilde{\faultindicator}}
\newcommand{\testweightindicator}{\widetilde{\weightindicator}}
\newcommand{\testfaultcontainer}{\widetilde{\faultcontainer}}

\makeatletter
\renewcommand\subsubsection{\@startsection{subsubsection}{3}{\z@}%
                       {-18\p@ \@plus -4\p@ \@minus -4\p@}%
                       {0.5em \@plus 0.22em \@minus 0.1em}%
                       {\normalfont\normalsize\bfseries\boldmath}}
\renewcommand\paragraph{\@startsection{paragraph}{4}{\z@}%
                       {-12\p@ \@plus -4\p@ \@minus -4\p@}%
                       {0.5em \@plus 0.22em \@minus 0.1em}%
                       {\normalfont\normalsize\itshape}}
\makeatother
\begin{document}
\title{\SMAUG: Secure Mobile Authentication\\Using Gestures}

\author{Christian~A.~Gorke
\and Frederik~Armknecht
}

\institute{University of Mannheim\\Mannheim, Germany\\
\email{gorke@uni-mannheim.de}, \email{armknecht@uni-mannheim.de}
}

\maketitle
\thispagestyle{plain}

%%%%%%%%%%%%%%%%%%%%%%%%%%%%%%%%%%%%%%%%%%%%%%%%%%%%%%%%%%%%%%%%%%%%

\begin{abstract}
We present \SMAUG\ (Secure Mobile Authentication Using Gestures), a novel biometric assisted authentication algorithm for mobile devices that is solely based on data collected from multiple sensors that are usually installed on modern devices -- touch screen, gyroscope and accelerometer.
As opposed to existing approaches, our system supports a fully flexible user input such as free-form gestures, multi-touch, and arbitrary amount of strokes.

Our experiments confirm that this approach provides a high level of robustness and security.
More precisely, in 77\% of all our test cases over all gestures considered, a user has been correctly identified during the first authentication attempt and in 99\% after the third attempt, while an attacker has been detected in 97\% of all test cases.
As an example, gestures that have a good balance between complexity and usability, e.g., drawing a two parallel lines using two fingers at the same time, 100\% success rate after three login attempts and 97\% impostor detection rate were given.
We stress that we consider the strongest possible attacker model: 
an attacker is not only allowed to monitor the legitimate user during the authentication process, but also receives additional information on the biometric properties, for example pressure, speed, rotation, and acceleration.
We see this method as a significant step beyond existing authentication methods that can be deployed directly to devices in use without the need of additional hardware.
\end{abstract}

\keywords{Mobile Security, Mobile Authentication, Authentication Schemes}

\section{Introduction}\label{sec:introduction}
Mobile devices such as mobile phones or tablets companion our everyday living and often store valuable personal data.
Consequently, there is a strong need for secure and usable authentication schemes.
The most popular approach is based on proving certain knowledge (``What I know''), i.e., the user has to insert a PIN or a pattern to gain access to the device.
However, this approach suffers from the known difficulties of choosing and memorizing strong passwords.
Moreover, it is subject to shoulder-surfing or smudge-attacks \cite{DBLP:conf/woot/AvivGMBS10,DBLP:conf/css/GaoJLL13}.
While this is true for any type of devices, these attacks are particularly dangerous for mobile devices where users have to frequently log-in while being outside of any secure environments like home or offices.
Even worse, often the device gives some visual feedback on the entered password, which makes these attacks even easier.
Consequently, alternative authentication mechanisms that are not based on knowledge (only) need to be considered.

A straightforward approach would be to include a second device (``What I possess''), e.g., some smart-card, but this negatively impacts usability.
A further alternative is to use biometric properties (``What I am'').
Apart from the fact that these methods often have their own security problems \cite{CCCTouchIdHack,ZWFingerprints,CCCIrisHack}, a further drawback is that this requires additional hardware, e.g., an iris scanner.
But any additional piece of hardware increases the size, the power consumption, and the price of the device, directly contradicting the market requirements.

\subsection{Contribution}
In this paper we present a novel biometric assisted authentication algorithm for mobile devices dubbed SMAUG (Secure Mobile Authentication Using Gestures) that overcomes all these limitations mentioned earlier.
The idea is based on the observation that modern mobile devices are equipped with a multitude of different types of sensors like the motion sensors gyroscope and accelerometer.
That is, users exhibit characteristic attributes like typing rhythm, gait, voice, and movement in general that can be measured with built-in sensors of a device and processed by an appropriate authentication algorithm.
For authentication based on \SMAUG, a user starts with registering a gesture (enrollment phase) and authenticates to the device later on by re-drawing the registered gesture (verification phase).
Here, a gesture is any interaction with the device regarding the touch screen or motion sensors, e.g. a drawn shape on the surface of the device, moving the device through space, or both simultaneously.
Note that during touching the device, its position and rotation is changed depending on the position where it is being touched by the user.
During both the enrollment and verification phase, a number of sensor data are collected.
In order to authenticate, our algorithm combines these information to enable \emph{multi-factor} authentication: a user needs to \emph{know} what gesture has to be made but also \emph{how} the gesture needs to be performed.
Hence, the user re-draws the earlier registered gesture and, if done by the legitimate user, he will be validated correctly with a high probability.
On the other hand, an impostor will be rejected by \SMAUG\ with a high probability.
Please note that \SMAUG\ supports multi-gesture login, that is the user may register more than only one gesture.
At the beginning of the verification phase \SMAUG\ randomly chooses a gesture out of this set of all registered gestures, but possibly hiding the name of the gesture.
Our scheme exhibits a number of remarkable features as follows:

\begin{description}
\item[No Dedicated Hardware Required:]
Our algorithm uses only hardware that is commonly found in mobile devices: a touch screen, an accelerometer, and a gyroscope.
This saves costs and makes \SMAUG\ accessible on a huge number of devices.
\item[Fully Flexible Input:]
As opposed to previous approaches, the user can choose \emph{any} arbitrary gesture: it does not need to be chosen from a predefined set, respects multi-touch, and can consist of as many strokes as desired at the same time.
Furthermore, any number of gestures may be registered.
In other words, \SMAUG\ does not limit form and length of the user's gesture.
\item[Efficient Feature Extraction and Detection:]
We explain in detail how collected data can be combined in order to extract over 320 features for each gesture.
Utilizing dynamic time warping allows us having efficient feature extraction and detection.
Furthermore, we introduce individualized weighting for each user, gesture, and even feature for a very adaptive and distinct verification process.
\SMAUG\ employs not only the features of single sensors, but fuses them in different ways.
The motion sensors are synchronized and interpreted as a single sensor, without excluding individual sensor features.
Data from the touch screen is structured in a way which allows for the extraction of stroke dependent features for each single stroke in addition to features applying to the whole gesture at once.
Merging everything together yields an entangled net of features allowing to distinguish between users.
Finally, by supporting seven system parameters and four security parameters we provide a high level of flexibility for adapting the behavior of the algorithm, depending on the system and policies, e.g., allowed retries after failed login attempts or sensitivity of the verification.
In addition, we specify default values to allow a start out of the box.
\item[High Security and Usability:]
We implemented a prototype of \SMAUG\ on Android 5.1.1 and tested it on a Nexus 5 device.
Note that \SMAUG\ is independent of operating system and device, as long as it supports touch screen, gyroscope, and accelerometer.
The results show that involving the efficient feature extraction and individual feature weighting pays off.
\SMAUG\ is capable to tell apart registered users from attackers with very high probability.
We stress that we consider the strongest possible attacker model:
an attacker was not only allowed to watch the legitimate user drawing the identifying gesture each time, but also receives additional information on the biometric properties, for example pressure, speed, rotation and acceleration.
We are not aware of any other authentication scheme based on this type of data that achieves a similar level of accuracy and security.
Surprisingly this strong level of security is already achieved even if the user has chosen a very simply figure like the letter ``A'' (denoted by gesture \gesture{A}).  

With respect to usability, the registration process remains modest.
A user is required to perform a gesture ten times to provide sufficient data to the algorithm for gesture registration.
This is one of the system parameters mentioned above and may be changed.
However, we found that ten is well chosen between time time investment and security.
Being the most time-consuming part, however, registration process needs to be done only once.
After the features are extracted and processed, the authentication process itself takes only a fraction of a second.
Also note that we do not employ continuous authentication, therefore our algorithm works offline and the privacy of the user is untouched as well as the battery life.
\end{description}
We also want to point out that biometric schemes which rely on direct biometric properties, such as iris scan or fingerprints, have a real-world application problem when it comes to leakage of this data.
In other words, if such a biometric property is leaked, it is not safe anymore, i.e., (partially) public known.
However, a user can change these only a very limited time, in the case of fingerprints only nine times.
Hence, we think a scheme like \SMAUG\ which is based on the features of biometric properties, and not the properties itself, is much more secure in the long term.
Removing old gestures and adding new ones is a matter of a few seconds and can be repeated arbitrarily often.
There is no limit besides the hardware storage and the user's imagination regarding the amount of gestures.
This feature is known from traditional authentication methods like passwords or certificates.

With respect to usability, a user may want to use \SMAUG\ with the same gestures on each of her devices.
However, since \SMAUG\ does not exchange information from any device to another, hence a user has to enroll the same gesture on each device.
If \SMAUG\ is not only used for device authentication but for service authentication on multiple devices, the same gesture may still be used.
This can be realized by storing an additional device identifier at the authentication service, which holds the user identification and gesture templates.

\subsection{Comparison to State of the Art}
Compared to existing work, our scheme \SMAUG\ provides the richest set of capabilities:
multi-gesture (arbitrary amount of gestures),
free-form gesture (gestures are not restricted in form or length),
multi-touch (arbitrary amount of fingers used at the same time),
multi-stroke (each finger can draw an arbitrary amount independent of other fingers),
multi-factor (combining biometric data with knowledge),
multi-sensor (touch sensor, gyroscope, accelerometer),
sensor fusion (combined evaluation of sensor data, especially gyroscope and accelerometer data),
and graphical password support (background images for gestures) at the same time.
Additionally, \SMAUG\ does not use continuous authentication and hence works offline, i.e., it runs on the device itself without the need to communicate to another party.
Furthermore, \SMAUG\ is independent of the operating system and hence can be implemented from most popular to rare systems.
Moreover, \SMAUG\ provides security with respect to the strongest attacker model considered so far, i.e., an attacker who has full knowledge on how the gesture looks like and how it has been drawn.

In Table~\ref{tbl:authschemes} on page~\pageref{tbl:authschemes} we give an overview and comparison of mobile authentication schemes as well as their capabilities.
Each row in this table refers to the work given in the first column.
A (red) ``\ctableno'' states a missing feature, while a (green) ``\ctableyes'' denotes that the feature is fully supported in the according work.
Other entries, like (yellow) ``partly''  or (yellow) ``yes'' denote a general support of a certain feature, but it is a restriction in comparison to other work, e.g., if only one operating system is supported.
Note that the colors of ``\ctableyes'' and ``\ctableno'' are interchanged for continuous authentication since it relates to loss of privacy, which we interpret as a drawback of such schemes.
We are not aware of schemes which run this operation efficiently on the device itself.
As the table shows, our algorithm \SMAUG\ is rich on features and steps ahead of existing work, providing the richest set of capabilities.

\subsection{Outline}
This paper is structured as follows.
In Section~\ref{sec:preliminaries}, we describe the basic components and the considered attacker model.
Section~\ref{sec:overview} provides an overview of the \SMAUG\ scheme which is divided into an enrollment phase and a verification phase.
These are explained in detail in Section~\ref{sec:enrollmentphase} and Section~\ref{sec:verificationphase}, respectively.
Section~\ref{sec:experiments} presents an implementation of our scheme and discusses various experimental results.
Section~\ref{sec:RelatedWork} gives an overview of related work and Section~\ref{sec:Conclusion} concludes the paper including open questions for future research.

\section{Preliminaries}\label{sec:preliminaries}
In this section we explain the basic components of our authentication algorithm \SMAUG\ and the considered attacker model.

\subsection{Sensors}\label{sec:sensors}
\SMAUG\ utilizes three types of sensors that are commonly found in mobile devices: touch screen, gyroscope, and accelerometer.
Touch screens report so-called \emph{touch events}, each consisting of tap coordinates, pressure, size of the pointer, and a \emph{touch event action}.
A \emph{pointer} is a general term for input devices used on the touch screen -- usually this is a finger or a pen-like device.
Modern mobile devices also support \emph{multi-touch} events, i.e. allowing more than one pointer at the same time on the touch screen.
When multiple pointers are used, each change of one pointer results in a new touch event of all active pointers.
We call this set of events a \emph{touch event set}.
If only one finger at the same time is being used on the touch screen overall, we denote this as \emph{single-touch}.
Hence, all other cases are called multi-touch.
Besides the touch sensor, the \emph{gyroscope} reports the change of the angular rate for all three space dimensions while an \emph{accelerometer} measures the acceleration in each space dimension.

In addition to the touch screen, we make use of two motion sensors, usually implemented in hardware as MEMS (microelectromechanical systems) sensors due to their size and energy consumption.
A \emph{gyroscope} reports the change of the angular rate for all three space dimensions while an \emph{accelerometer} measures the acceleration in each space dimension.
Every modern operating system supporting accelerometers also provides a \emph{linear acceleration} value which removes earth's gravitation in the measurement.
In this paper, when we write accelerometer, we always refer to the linear one.
Important for our authentication algorithm is the frequency of these sensors which determines the accuracy of the measured values. 
While they work with very high precision, the frequency is usually limited by the operating system to avoid an inflow of too many data from the sensors. 
Thus, we will operate on the highest frequency that is allowed by the operating system, but at most 200Hz.
For the touch screen, this is usually 30Hz or 60Hz, which is sufficient for \SMAUG.
Each measurement of the motion sensors is called a \emph{motion event}.
\SMAUG\ considers not only the measurements of each motion sensor, but also combines the motion sensor values into a new set, what we denote by \emph{motion fusion}.
This is explained in detail in Section~ \ref{sec:motionfusionfeatures}.

\subsection{Gestures}\label{sec:gestures}
We denote a simultaneous sequence of touch and motion events as a \emph{gesture}.
Each gesture in \SMAUG\ consists of an arbitrary number of strokes.
A \emph{stroke} is a sequence of touch events belonging to the same pointer.
This sequence starts with a down event (touching the screen) and ends eventually with an up event (releasing the pointer from the screen).
For our authentication algorithm we require at least one stroke to be part of a gesture.
A stroke with drawing length zero is called a \emph{point}.
In comparison to strokes, a point has a time length but no distance.
For our authentication algorithm we also allow \emph{multi-strokes}, meaning a gesture can consist of an unlimited amount of simultaneous strokes.
Observe that this may take place at overlapping time periods, that is one stroke may end and a new one starts while another is still active.
Furthermore, between two strokes it is not required to stay in contact with the touch screen.
We call a \emph{stroke gap} the time period between strokes where no touch event happens.
However, note that within stroke gaps, motion data events still occur.
A \emph{stroke run} is a set of strokes separated by stroke gaps. Finally, we define a \emph{closed gesture} to be a gesture without stroke gaps.
\SMAUG\ supports all theses features to allow for very individual inputs.
This covers a wide range of gestures, from drawings and signatures, to letters, rhythms, and even movements of the device in space due to the combination of touch and motion sensors.

\subsection{Dynamic Time Warping}\label{sec:dtw}
One core aspect of authentication algorithms based on gestures is that a gesture of a user needs to be mapped to already stored data such that later on, this gesture will be recognized if coming from the same user.
Here, one has to cope with two contradicting requirements, security and robustness, which excludes a number of existing techniques as we elaborate now.

Due to the security requirements, it is important that the order of (overlaying) strokes is registered.
Thus, image recognition algorithms cannot be used here since strokes may overlap, e.g. due to multi-stroke and multi-touch.
This holds in particular for algorithms that verify handwritten signatures.

An alternative seems to be machine learning algorithms.
However, the class of supervised algorithms learns a classification rule by getting valid and invalid examples, e.g, Support Vector Machines or $k$-nearest neighbor.
In this case, the valid examples would be the gestures executed by the legitimate user.
However, it remains unclear how and which invalid examples should be generated.
For example, if they stem from different persons, they will significantly different from the target gesture with high probability which negatively impacts the quality of the learning result.
A way out seems to be to limit the possible gestures to a given set of gestures which have to be used by every user.
Then the gestures can be compared to each other.
However, this would contradict our goal that we want to offer the maximum flexibility (and entropy) when choosing the gesture.
On the other hand, unsupervised algorithms will either be too generous, that is very robust but not secure, or need too much computing power which is not provided locally on mobile devices.

An even more natural choice would be to use one of the already existing gesture recognition algorithms, for example \$1 and \$N which cover single-touch multi-stroke cases \cite{DBLP:conf/uist/WobbrockWL07,DBLP:conf/graphicsinterface/AnthonyW12}.
However, they ignore the order of the strokes, may add additional strokes to reduce multiple strokes back to a single one, and are in general too ``forgiving'' for bad inputs, e.g., by resizing and rotating inputs.
This again increases robustness drastically, but decreases the level of security.

Given these considerations, we opted for the dynamic time warping algorithm (DTW), originating from speech recognition \cite{Sakoe:1990:DPA:108235.108244}.
DTW can be used to measure the distance between two sets, or the similarity between two sequences for real values -- being $0$ for equal sets.
DTW comes with the property that the sequences don't need to have the same number of elements and don't have to be normalized.
For example, the values of the $x$-coordinates of one stroke can be compared directly to the $x$-values of another stroke.
The result is a measurement of their distance, being $0$ if both sets are equal.
The data extracted by \SMAUG\ of a gesture contains plenty of these temporal sequences and is therefore well suited for DTW. 
Plus, this technique was already used in earlier work to recognize biometrical inputs \cite{DBLP:conf/chi/LucaHBLH12,MTroMultiFact}.

We will now briefly define the DTW algorithm.
Let $N := \{N_1, \ldots, N_n\} \in \R^n$, $M := \{M_1, \ldots , M_m\} \in \R^m$.
We set $\delta(a, b) := \left|a-b\right|$ as the distance function for $(a,b) \in (N,M)$.
The dynamic time warping algorithm $\DTW(N, M)$ is then defined as given in Algorithm~\ref{alg:dtw}, where $[n] := \{0,\ldots,n\}$.

\begin{algorithm}[htb]
\DontPrintSemicolon
\KwIn{$N = (N_1, \ldots, N_n)\in \R^n$, $M = (M_1, \ldots , M_m)\in\R^m$}
\KwOut{$\DTW(N, M) := \delta(n,m)$}
$\delta(i,j) \gets 0 ~\forall~ i\in[n], j\in[m]$\;
\For{$i \leftarrow 1$ \KwTo $n$}{$\delta(i,0) \gets \infty$}
\For{$j \leftarrow 1$ \KwTo $m$}{$\delta(0,j) \gets \infty$}
\For{$i \leftarrow 1$ \KwTo $n$}{
\For{$j \leftarrow 1$ \KwTo $m$}{
$\delta(i,j) \gets \left|N_i- M_j\right| + \min\left(\delta(i-1,j), \delta(i,j-1), \delta(i-1,j-1)\right)$
}
}
\KwRet{$\delta(n,m)$}\;
\caption{Dynamic Time Warping (DTW).}
\label{alg:dtw}
\end{algorithm}

\subsection{Adversary Model}\label{sec:adversarymodel}
Our goal is to provide protection against the strongest possible attacker model.
Therefore, we consider an adversary who is allowed to observe the whole enrollment phase.
This includes direct vision to the touch screen of the device and additionally a brief explanation on how the input has been done.
Clearly, this is stronger than shoulder-surfing or a smudge attack, since the adversary gets all of this information for free.
One remarkable property of \SMAUG\ is that indeed security against this type of attacker is given.

\section{Algorithm Overview}\label{sec:overview}
In this section we give an overview of \SMAUG, our proposed biometric authentication algorithm for mobile devices.
This includes notation, algorithm work flow, description of the algorithm modes, and an overview of all system and security parameters.

\subsection{Notation}
In a nutshell, the algorithm operates on two categories of data: meta data and sensor data.
The data belongs to a certain gesture $\anygesture$, we will employ $\anygesture$ to refer to an arbitrary gesture.
The meta data set contains general information and will be denoted by $\metadata$.
The sensor data set comprises three types of measurements: touch data $\touchdata{0}$, accelerometer data $\laccdata{0}$, and gyroscope data $\gyrodata{0}$.
Furthermore, fusion data $\fusiondata{0}$ will be computed from both motion sensor data sets. 
The data sets will be processed and altered by different algorithms. 
New results of computation on these data sets will be denoted by increasing superscript indexes, i.e., $\touchdata{0},\touchdata{1},\touchdata{2},\ldots$ and so on.

The sensor data sets have a similar structure.
Each set is represented by a matrix over real values where a row represents one data event during the measurements and each column represents one property of the data.
For example, each row of $\touchdata{0}$ represents one touch event which by itself is a vector where each entry, i.e., column, represents one property, for example the pointer's pressure.
When we refer to a specific event, i.e., row, we use the notation $\touchdata{k}_i$ to address the $i$-th touch event of $\touchdata{k}$.
Moreover, if $x$ refers to a certain property, we denote by $\touchdata{k}[x]$ the column that stands for property $x$
To refer to the value of a property $x$ within the $i$-th event, we write $\touchdata{k}_i[x]$.
We use analogous notation with respect to the accelerometer data $\laccdata{k}$ and gyroscope data $\gyrodata{k}$.
To keep the description short, we use $\motiondata{k}$ to refer to the data sets $\gyrodata{k}$, $\laccdata{k}$, and $\fusiondata{k}$ at the same time. 
By $\left(\touchdata{k}\right)_\round$ we refer to the data set of round $\round$, we omit this subscript if $\round$ is clear from the context, for example inside an algorithm.
Here, a round describes how often a user has to enter the same gesture during the enrollment phase in order to correctly register it.

By $\user$ we denote the user, who is the owner of the device and the input data.
An unknown user and possible impostor is denoted by $\unknownuser$.

\subsection{Algorithm Work Flow}
In a nutshell, \SMAUG\ runs on mobile devices and consists of an enrollment phase for each gesture a user wants to use for authentication.
\SMAUG\ extracts specific features of this data which represent the knowledge and behavior of the user during login.
This data is personalized through certain feature weights which depend on the form of the gesture and the user at the same time.
It is very important to note that the selected features of the data as well as the architecture of \SMAUG\ allow for a combination of arbitrarily inputs, such as free-form, multi-touch, multi-strokes, arbitrary number and length of strokes, resulting in a flexible input system only limited by the user herself.
As a second part, the verification phase asks the user to enter a randomly chosen gesture out of all gestures entered during the enrollment phase and records a single gesture input.
Then, the same certain data features are extracted as before and compared to the previously stored authentication gesture.
If the verification or authentication input consists of the same data features as the corresponding one from the enrollment phase, the user is authenticated.
However, due to the personalized weights and feature handling, an impostor is not able to authenticate with high probability.

\begin{figure}[!thb]
\centering
\begin{tikzpicture}[
  every node/.style={
    rectangle,
    draw=none,
    %very thick,
    minimum height=3em,
    minimum width=10em,
    inner sep=2pt,
    align=center,
    scale=0.8,
    transform shape,
    node distance=2em,
  },
  phase node/.style={
  	node distance=0em,
  	minimum height=2.5em,
  },
  server color/.style={
  	fill=green!30,
%  	postaction={
%		pattern=dots,
%  		pattern color=black!99!green!30,
%  		draw=black,
%  	},
  },
  user color/.style={
  	fill=red!30,
%  	postaction={
%  		pattern=north west lines,
%  		pattern color=black!99!red!30,
%  		draw=black,
%  	},
  },
  server node/.style={
  	server color,
  	draw=black,
  },
  user node/.style={
  	user color,
  	rounded corners=5pt,
  	draw=black,
  },
  leftarrownode/.style={
  	left,
  	midway,
  	xshift=-0.5em,
  	scale=0.9,
  	align=left,
  	inner sep=0pt,
  	minimum width=1em,
  },
  rightarrownode/.style={
  	right,
  	midway,
  	xshift=0.5em,
  	scale=0.9,
  	align=left,
  	inner sep=0pt,
  	minimum width=1em,
  },
  user arrow/.style={
  	double -latex=1pt colored by black and red!30,
  },
  server arrow/.style={
  	double -latex=1pt colored by black and green!30,
  },
  >=latex, %Make the arrow tips latex
  myline/.style={draw, very thick,black, node distance=1.1cm},
  shorter/.style={shorten <=0mm,shorten >=0.1mm},
  node distance=1.5em,
]

\begin{scope}[every node/.append style={phase node}]
\node[label={[shift={(-2.5em,0em)}]center:\textbf{Enrollment Phase}}] (EP) {};
\end{scope}

\begin{scope}[every node/.append style={user node}]
\node[below=of EP, yshift=1em] (CG) {Choose Gesture};
\node[below=of CG] (IG) {Input Gesture};
\end{scope}

\begin{scope}[every node/.append style={server node}]
\node[right=of CG, xshift=2em] (IP) {Input Prompt};
\end{scope}

\begin{scope}[every node/.append style={user node}]
\node[below=of IP] (IG2) {Input Gesture};
\end{scope}

\begin{scope}[every node/.append style={server node}]
\node[below=of IG] (PP) {Postprocessing};
\node[below=of PP] (FE) {Feature Extraction};
\node[below=of FE] (T) {Template Database};
\node[below=of T] (W) {Weights Database};
\node[below=of W] (TH) {Thresholds};
\node[below=of IG2] (PP2) {Postprocessing};
\node[below=of PP2] (FE2) {Feature Extraction};
\end{scope}

\node[draw=none] at ($(T)!0.5!(W)$) (PH) {~}; %Placeholder to get correct spacing

\begin{scope}[every node/.append style={server node}]
\node at (PH -| FE2) (V) {Validation};
\node at (TH -| V) (Ver) {Verification};
\end{scope}

\begin{scope}[every node/.append style={user node}]
\node[ellipse, right=of Ver, xshift=2em] (AS) {Authentication\\Success};
\node[ellipse, above=of AS] (AF) {Authentication\\Failure};
\end{scope}

\node[draw=none] at ($(Ver)!0.5!(AS)$) (PH2) {~}; %Placeholder to get correct spacing

\begin{scope}[every node/.append style={phase node}]
\node at (EP -| PH2) (VP) {\textbf{Verification Phase}};
\end{scope}

\draw[user arrow] (CG.south)--(IG.north);
\draw[->] (IG) edge [loop left, out=-175, in=175, looseness=5] node[minimum width=5em, scale=0.9, inner sep=0pt] {repeat\\$\sprounds$ times} (IG);
\draw[server arrow] (IG.south)-- node [leftarrownode] {Fetch} node[rightarrownode] {Data} (PP.north);
\draw[server arrow] (PP.south)-- node [leftarrownode] {Process} node[rightarrownode] {Data} (FE.north);
\draw[server arrow] (FE.south)-- node [leftarrownode] {Generate} node[rightarrownode] {Template} (T.north);
\draw[server arrow] (T.south)-- node [leftarrownode] {Compute} node[rightarrownode] {Weights} (W.north);
\draw[server arrow] (W.south)-- node [leftarrownode] {Compute} node[rightarrownode] {Thresholds} (TH.north);
\draw[server arrow] (IP.south)--(IG2.north);
\draw[server arrow] (IG2.south)-- node [leftarrownode] {Fetch} node[rightarrownode] {Data} (PP2.north);
\draw[server arrow] (PP2.south)-- node [leftarrownode] {Process} node[rightarrownode] {Data} (FE2.north);
\draw[server arrow] (FE2.south)-- node [leftarrownode] {Compute} node[rightarrownode] {Pre-Template} (V.north);
\draw[server arrow] (V.south)-- node [leftarrownode] {Fetch} node[rightarrownode] {Result} (Ver.north);
\draw[server arrow] (Ver.north east)-- node [above, midway, scale=1.1, yshift=-1em, xshift=-0.8em] {\xmark} (AF.west);
\draw[server arrow] (Ver.east)-- node [below, midway, scale=1.1, yshift=0.5em] {\cmark} (AS.west);
\draw[user arrow] (AF.north) -- node[above, sloped, midway, scale=0.9, xshift=-0.5em, yshift=-0.5em] {repeat at most $\sprepeat$ number of times} (IP.south east);

\draw[server arrow] (T.east)--([yshift=0.5em]V.west);
\draw[server arrow] (W.east)--([yshift=-0.5em]V.west);
\draw[server arrow] (TH.east)--(Ver.west);

\draw[dashed] ([xshift=-5em]EP.south west) -- ([xshift=-5em, yshift=-1.3em]TH.south west) ([xshift=-5em, yshift=-1.3em]TH.south west) -- ([xshift=1em, yshift=-1.3em]TH.south east) ([xshift=1em, yshift=-1.3em]TH.south east) -- ([xshift=1em]EP.south east) ([xshift=1em]EP.south east) -- ([xshift=-5em]EP.south west);
\draw[dashed] ([xshift=-7em]VP.south west) -- ([xshift=-1.1em, yshift=-1.3em]Ver.south west) ([xshift=-1.1em, yshift=-1.3em]Ver.south west) -- ([xshift=1.7em, yshift=-1.4em]AS.south east) ([xshift=1.7em, yshift=-1.4em]AS.south east) -- ([xshift=6.5em]VP.south east) ([xshift=6.5em]VP.south east) -- ([xshift=-7em]VP.south west);
\end{tikzpicture}
\caption{Schematic flow chart of the enrollment phase and verification phase of \SMAUG.
The steps displayed as rounded boxes (red) are done by involvement of the user, the rectangles (green) part by \SMAUG.
This respectively holds for the arrows.
The value $\sprounds$ is a system parameter and can be changed, default value is $10$.
Also, the value of $\sprepeat$ is free of choice and by default $2$.
At the end, verification outputs a binary decision (to the user): authentication success (\cmark) or authentication failure (\xmark).}
\label{fig:flow}
\end{figure}
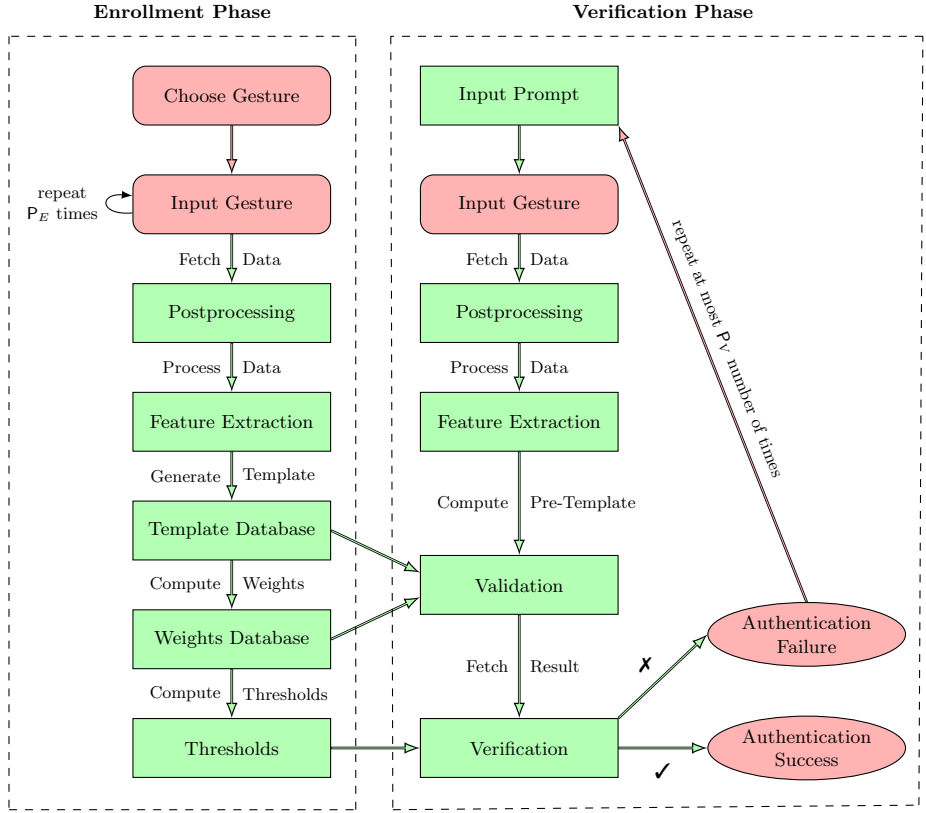

As already mentioned, our proposed scheme consists of two different phases: an enrollment phase where the user $\user$ registers a gesture $\anygesture$, and a verification phase where new input is compared to the registered gesture $\anygesture$.
An overview of the algorithm work flow is depicted in Fig.~\ref{fig:flow}.
This schematic shows the enrollment phase and verification phase of \SMAUG, where the rounded boxes (red) are performed by the user, and the rectangular boxes (green) are done by \SMAUG.
At least one enrollment phase (gesture registration) must be executed before a verification phase (authentication) can take place.
The output is either an authentication failure or success which is output to the user, e.g. by device unlocking, online service authentication, and so on.
At the end of this section we also give a brief note on the real-world instantiation of \SMAUG.

\subsubsection{Enrollment Phase}\label{sec:enrollmentflow}
First, in the \emph{data processing stage}, the user inputs her gesture $\anygesture$ overall $\sprounds$ times, yielding the data sets $(\metadata, \touchdata{0}, \gyrodata{0}, \laccdata{0})_\round$ for each round $\round=1,\ldots,\sprounds$.
Then, each data set gets processed and features of the data, such as coordinates and other properties, are extracted for each round $\round$.
Finally, the templates for this gesture are generated and stored in a template database, containing all relevant information for each gesture.

Second, in the \emph{weight computation stage}, we do a cross validation of the user data.
That is, for each round $\round$ the appropriate data set from the data processing stage is being treated as a single test input and compared to the already stored templates of the gesture.
We then count the number of erroneous deviations for each feature and compute the weight of each feature depending on the number of errors for this specific feature.
The weights represent the ``stability'' or ``habitages'' of the user mapped onto the feature set.
For example, if a feature failed in nearly all $\sprounds$ tests, it will have a very low weight.
But if a feature in all tests always succeeds, it will have the maximum weight, therefore being very important  for this gesture and user.
Here, feature failure relates to high deviation from all other data sets regarding this feature compared to a specific data set.
In the same way, success refers to a low deviation.
Furthermore, each feature belongs to one of three classes, called \emph{tiers} of features.
The weight also depends on the value of this tier, reflecting the influence of certain features.
For example, if the $x$ and $y$ coordinates of the touch sensor are horribly wrong, but the Kurtosis of one axis of a motion sensor is perfect, the first must weight in more.
However, the second is important for our strong attacker model.
The set of all weights of the gesture and its strokes is denoted by $\weightsset$.
Finally, out of $\weightsset$ two thresholds $\threshold_1, \threshold_2$ get computed.

The inputs and outputs of these subsidiary algorithms (or stages) as well as their order is shown in the algorithmic description of Algorithm~\ref{alg:enrollment}.
%Here, $\algFeatureExtractionShort$ and $\algTemplateGenerationShort$ are abbreviations for $\algFeatureExtraction$ and $\algTemplateGeneration$, respectively.
Strictly speaking, the for loop in Line~1 does not necessarily mean that all three algorithms have to be performed, data gathering may be done completely separated from the rest.
The enrollment phase of \SMAUG\ is explained in detail in Section~\ref{sec:enrollmentphase}.

\begin{algorithm}[htb]
\DontPrintSemicolon
\KwIn{User $\user$, gesture $\anygesture$}
\KwOut{$\metadataset, \touchdataset, \motiondataset, \weightsset, \threshold_1, \threshold_2$}
\For{$\round \leftarrow 1$ \KwTo $\sprounds$}{
$(\metadata, \touchdata{0}, \gyrodata{0}, \laccdata{0})_\round \gets \algDataGathering(\user, \anygesture, \round)$\;
$(\touchdata{1}, \motiondata{1})_\round \gets \algPostProcessing((\touchdata{0}, \gyrodata{0}, \laccdata{0})_\round)$\;
$(\touchdata{1}, \touchdata{2}, \touchdata{3}, \motiondata{2})_\round \gets \algFeatureExtraction((\touchdata{1}, \motiondata{1})_\round)$\;
}
$(\touchdataset,  \motiondataset) \gets \algTemplateGeneration(\metadata, \{(\touchdata{1}, \touchdata{2}, \touchdata{3}, \fusiondata{1}, \motiondata{2})_r\}_{1\leq \round \leq \sprounds})$\;
$\weightsset \gets \algComputeWeights(\touchdataset, \motiondataset)$\;
$(\threshold_1,\threshold_2) \gets \algThreshold(\weightsset)$\;
\KwRet{$\metadataset, \touchdataset, \motiondataset, \weightsset, \threshold_1, \threshold_2$}\;
\caption{Enrollment phase of \SMAUG.}
\label{alg:enrollment}
\end{algorithm}

\subsubsection{Verification Phase}\label{sec:verificationflow}
The verification phase is executed each time an authentication attempt takes place. 
This phase is divided into two stages which are repeated until a successful input was made, but at most $\sprepeat$ times.

First, in the data is gathered processed, that is an unknown user $\unknownuser$ enters a single data set $(\testmetadata, \testtouchdata{0}, \testgyrodata{0}, \testlaccdata{0})$ by performing a gesture $\anyunknowngesture$.
This data set is being processed and features are extracted.

Second, to perform the verification, the data set and features from the previous stage are compared to the data set and features belonging to the original gesture template $(\touchdataset, \motiondataset)$ which are kept in the template database.
We count the number of new errors for each feature of this comparison as $\faultscount$.
Here, an error is defined as too much deviation from the stored gesture template.
We also include all weights of the errors from $\weightsset$, yielding the sum $\faultsweight$ over all weighted errors.
If $\faultsweight \leq \threshold_1$ and $\faultscount \leq \threshold_2$, the algorithm returns the decision \True\ and the authentication is successful, else \False.

Algorithm~\ref{alg:verification} gives an algorithmic description of this phase.
The verification algorithm computes a binary decision $\decision$ which denotes if the authentication attempt was successful.
We explain the verification phase of \SMAUG\ in detail in Section~\ref{sec:verificationphase}.

\begin{algorithm}[htb]
\DontPrintSemicolon
\KwIn{$\touchdataset, \motiondataset, \weightsset, \threshold_1, \threshold_2, \unknownuser, \anyunknowngesture$}
\KwOut{Decision $\decision$}
$\decision \gets \False$\;
$\authtry \gets 0$\;
\While{$\authtry \leq \sprepeat$}{
$(\testmetadata, \testtouchdata{0}, \testgyrodata{0}, \testlaccdata{0}) \gets \algDataGathering(\unknownuser, \anyunknowngesture, \authtry)$\;
$(\testtouchdata{1}, \testmotiondata{1}) \gets \algPostProcessing(\testtouchdata{0}, \testgyrodata{0}, \testlaccdata{0})$\;
$(\testtouchdata{2}, \testmotiondata{2}) \gets \algTestFeatureExtraction(\testtouchdata{1}, \testmotiondata{1})$\;
$(\testweightindicator, \testfaultindicator) \gets \algValidation((\testtouchdata{2}, \testmotiondata{2}), (\touchdataset, \motiondataset), \weightsset)$\;
\If{$\algVerification(\faultsweight, \faultscount, \threshold_1, \threshold_2) = \True$}{
$\decision \gets \True$\;
$\authtry \gets \sprepeat$\;
}
$\authtry \gets \authtry+1$\;
}
\KwRet{$\decision$}\;
\caption{Verification phase of \SMAUG.}
\label{alg:verification}
\end{algorithm}

\subsection{Algorithm Modes}\label{sec:modes}
\SMAUG\ supports three different modes which can be combined in any way.

\begin{description}
\item[Background Image Mode.]
This mode is enabled whenever the owner of the gesture chooses to add a background image to the gesture during the enrollment phase.
This picture may give helping indicators for how and especially where to perform the gesture on the touch screen.
As shown in the results of our experiments, cf. Section~\ref{sec:results}, this does not impact the security of \SMAUG\ since we are using a different set of security parameters for this mode.
\item[Multi-Touch Mode.]
The algorithm computes the number of maximum pointers at the same time for each gesture during the feature extraction phase, cf. Section~\ref{sec:touchroundfeatures}.
Therefore this mode is chosen automatically dependent on the gesture data.
For this mode we also use a different set of security parameters, which are explained in Section~\ref{sec:threshold}.
\item[Secret Gesture Mode]
If this mode is disabled, during the login on the device a message will be displayed instructing the user which gesture to enter by displaying the name of the gesture.
If this mode is enabled, this message is hidden.
By default, this mode is enabled, hence \SMAUG\ performs multi-factor authentication.
If the user has created multiple gestures and flagged as secret mode, she must assure that each of them has different backgrounds to distinguish between them.
We do not employ specific security parameters for this mode, since our adversary model covers a well-informed attacker, no matter of this mode.
\end{description}

\subsection{System Parameters}\label{sec:parameters}
We summarize all seven system parameters and the secret gesture mode in Table~\ref{tbl:parameters}.
For each parameter, a default value is given as well as a brief description.
The values of the security parameters are given in Section~\ref{sec:thresholds}, Table~\ref{tbl:secparams}.

\begin{table}[htb]
\centering
\begin{tabular}{ccp{60mm}}
\toprule
\textbf{System Parameter} & \textbf{Default Value} & \textbf{Description}\\
\midrule
\sprounds & $10$ & Number of times a user has to repeat a gesture during enrollment phase, cf. Section~\ref{sec:enrollmentphase}).\\
\sprepeat & $2$ & Maximum number of additional authentication attempts during verification phase, cf. Section~\ref{sec:verificationphase}.\\
\sptierone & $0.75$ & Tier one feature weight multiplicator, cf. Section~\ref{sec:algweightcalculation}.\\
\sptiertwo & $1$ & Tier two feature weight multiplicator, cf. Section~\ref{sec:algweightcalculation}.\\
\sptierthree & $2$ & Tier three feature weight multiplicator, cf. Section~\ref{sec:algweightcalculation}.\\
\spmotionstartoffset & $150$ & Time offset in milliseconds for motion data before first touch event, cf. Section~\ref{sec:snuggle}.\\
\spmotionendoffset & $100$ & Time offset in milliseconds for motion data after last touch event, cf. Section~\ref{sec:snuggle}.\\
%\midrule
Secret Gesture Mode & Enabled & Display gesture name during verification phase, cf. Section~\ref{sec:modes}.\\
%\midrule
%\spfaultweightmult & \textit{cf. Table~\ref{tbl:secparams}} & Fault weight sum multiplicative security parameter, cf. Section~\ref{sec:threshold}.\\
%\spfaultweightadd & \textit{cf. Table~\ref{tbl:secparams}} & Fault weight sum additive security parameter, cf. Section~\ref{sec:threshold}.\\
%\spfaultnumbermult & \textit{cf. Table~\ref{tbl:secparams}} & Number of faults multiplicative security parameter, cf. Section~\ref{sec:threshold}.\\
%\spfaultnumberadd & \textit{cf. Table~\ref{tbl:secparams}} & Number of faults additive security parameter, cf. Section~\ref{sec:threshold}.\\
\bottomrule
\end{tabular}
\vspace*{2mm}
\caption{Default values and description of all system and security parameters.}
\label{tbl:parameters}
\end{table}

\subsection{Notes on Real-World Instantiation of \SMAUG}\label{sec:instantiationnote}
The description of both the enrollment phase and verification phase above, shows how \SMAUG\ works for a single gesture.
However, an instantiation of \SMAUG\ in practice will work slightly different.
As mentioned in the introduction, \SMAUG\ supports multiple gestures at the same time.
That is, a user can have as many gestures as she wants, as long as they all have different identifiers, i.e., descriptive labels such that the system can distinguish them like ``$\gesture{A}$''.
This is stored in the metadata employed in the enrollment phase.
Therefore, the enrollment phase of \SMAUG\ takes as input the user $\user$ and a set of gestures $\anygesture_1, \anygesture_2, \ldots$, which can be increased at any time.
Then each single gesture registration is performed as described in Algorithm~\ref{alg:enrollment}.
Note that the algorithm modes are also gesture dependent.

For the verification phase, a real-world-instantiation of \SMAUG\ will randomly choose one of all registered gestures of the user.
This is part of the input prompt, cf. Figure~\ref{fig:flow}.
Furthermore, if the authentication fails more than $\sprepeat+1$ times, a fallback mechanisms will be called.
This mechanism is out of scope for this paper, but may be implemented as a password authentication, or increasing the time between authentication trials.

In the remainder of this paper we will describe the algorithms for one gesture, if not stated otherwise.
An extension merely adds another index to most variables.

\section{Enrollment Phase}\label{sec:enrollmentphase}
The enrollment phase is repeated $\sprounds$ times for a gesture $\anygesture$ and each repetition is called a \emph{round} $\round$.
According to our experiments, choosing  $\sprounds=10$ achieves the highest amount of security while still providing a reasonable level of usability.
Before the first round, the user chooses the name and mode of the gesture.
At the end of this phase, the device stores a kind of fingerprint of the gesture defined and personalized by the user who entered the gesture in form of a template and according weights.
Note that the user can create as many gestures as she wants to.
However, for the sake of readability we focus on one gesture within this section.

\subsection{Data Gathering}\label{sec:datagathering}
The first step is having the user $\user$ choosing the mode and entering the gesture $\anygesture$ for each round $\round$.
The algorithm $\algDataGathering$ records the following data sets during this phase:
\begin{align*}
\left(\metadata, \touchdata{0}, \gyrodata{0}, \laccdata{0}\right)_\round \gets \algDataGathering(\user, \anygesture, \round).
\end{align*}

\begin{description}
\item[Meta Data $\metadata$:]
This includes the name of the gesture, a unique gesture identifier, the background image, the modes, and the round $\round$.
\item[Touch Data $\touchdata{0}$:]
When the user touches the touch screen, touch events get recorded and stored in $\touchdata{0}$.
The user is allowed to have stroke gaps in his gesture.
A single touch data event is a vector and contains a unique gesture identifier, the round number $\round$, and the following properties:
sensor type (touch screen),
event time (timestamp $\systemtimenano \in \N_0$ in nanoseconds),
pointer ID (identifier for a pointer, starting at $0$; for a stroke the pointer identifier always stays the same),
pointer number (counter starting at $0$ and counting up until the number of pointers in a single touch event set),
touch action (action of the touch event),
x- and y-coordinate $x, y \in \N_0$ of the pointer on the touch screen,
pressure $\pressure\in[0,2]$ (normalized pressure of the pointer on the touch screen; note that this is called pressure but measures also the size of a pointer on the touch screen),
size $\size\in[0,1]$ (normalized size of the pointer on the touch screen).

These values represent the maximum amount of physical information the device gets from a pointer on currently available touch screens.
The coordinates describe the layout of the gesture directly, while the pressure, size and system time values represent strong biometric identifier.
Recording is done with the highest frequency possible, for example 30Hz.
\item[Motion Data $\laccdata{0}$, $\gyrodata{0}$:]
The recording of motion data, i.e., gyroscope and accelerometer data, starts immediately after the user begins with each round and ends on finishing each round.
The user notifies the system when a round has finished by pressing a button on the screen.
Recording is done with the highest frequency possible, for example 200Hz.

As mentioned before, a motion data record ($\laccdata{0}$ and $\gyrodata{0}$) is a matrix where each row consists of the values of a single motion data event.
Each such event contains a unique gesture identifier, the round number $\round$, and the following properties:
sensor type (gyroscope or accelerometer),
event time (timestamp $\systemtimenano \in \N_0$ in nanoseconds),
three sensor values reported directly from the sensor.
While these raw values are already quite characteristic for a user we will describe the characteristic features below which can be computed from the sensor values.
\end{description}

\subsection{Post Processing}\label{sec:postprocessing}
The post processing algorithm corrects and selects the significant gesture data from the data gathered in Section~\ref{sec:datagathering} for each round $\round$.
It also computes the motion fusion data.
We denote this by
\[ (\touchdata{1}, \gyrodata{1}, \laccdata{1}, \fusiondata{1})_\round \gets \algPostProcessing((\touchdata{0}, \gyrodata{0}, \laccdata{0})_\round) \]
and explain the sub-algorithms of the post processing in the following.
Remember that $\motiondata{1} = (\gyrodata{1}, \laccdata{1}, \fusiondata{1})$.

\subsubsection{Touch Event Action Correction}\label{sec:touchactioncorrect}
The purpose of this first algorithm is to correct touch events in the data set which have an incorrect touch event action for our purpose.
These incorrect measurements appear due to the operating system's reporting of multiple touch events happen at the same time, i.e., releasing a finger from the screen while moving another one.
This is needed to identify strokes of the gesture later on.
In short, this procedure scans for all pointers that are still active after the \TouchPUp\ event and corrects these to a \TouchMove\ event.
The \TouchPUp\ event is triggered when more than one pointer is active on the touch screen and one of them is being released.
The event \TouchMove\ denotes motion of a single pointer.
We refer to Table~\ref{tbl:touchupcorrect} for an example of this process.
All changes are performed and stored in $\touchdata{0}$.
Remember that pointers are a general term for input object, such as finger or pen, see Section~\ref{sec:sensors}.

\begin{table}[htb]
\centering
\begin{tabular}{ccccc}
\toprule
\textbf{Event Time} & \textbf{Event Set} & \textbf{Pointer ID} & \textbf{Pointer Number} & \textbf{Touch Action}\\
\midrule
$\ldots$ & $\ldots$ & $\ldots$ & $\ldots$ & $\ldots$\\
208 & 5 & 0 & 0 & \TouchPUp\\
209 & 5 & 1 & 1 & \TouchPUp\\
210 & 5 & 2 & 2 & \TouchPUp\\
211 & 6 & 1 & 0 & \TouchMove\\
212 & 6 & 2 & 1 & \TouchMove\\
$\ldots$ & $\ldots$ & $\ldots$ & $\ldots$ & $\ldots$\\
\midrule
\multicolumn{5}{c}{$\Downarrow$ correction $\Downarrow$}\\
\midrule
$\ldots$ & $\ldots$ & $\ldots$ & $\ldots$ & $\ldots$\\
208 & 5 & 0 & 0 & \TouchPUp\\
209 & 5 & 1 & 1 & \TouchMove\\
210 & 5 & 2 & 2 & \TouchMove\\
211 & 6 & 1 & 0 & \TouchMove\\
212 & 6 & 2 & 1 & \TouchMove\\
$\ldots$ & $\ldots$ & $\ldots$ & $\ldots$ & $\ldots$\\
\bottomrule
\end{tabular}
\vspace*{2mm}
\caption{Example for \TouchPUp\ correction for two event sets in Android.
Without the correction, we would not be able to recognize which pointer left the set of active pointers on the touch screen.
The problem also exists for pointer joining, but this can be fixed during data gathering.
All events are sorted by event time and an event set contains all active pointers during a single measurement.
The pointer ID can be interpreted as a stroke identifier, and the pointer number just counts from zero up through all active pointers.
Touch action contains the touch event reported from the touch screen and operating system.}
\label{tbl:touchupcorrect}
\end{table}

\subsubsection{Strokes Determination}\label{sec:determinestrokes}
Observe that neither $\touchdata{0}$ nor $\motiondata{0}$ contains information about strokes.
Due to the previous correction, we now are able to detect and distinguish strokes.
Since each pointer has now a unique start and end event this algorithm computes unique identifiers for each stroke and each round $\round$, given the output of the previous algorithm. 
This allows us to compare strokes later on directly.

In a nutshell, for the identification we select all events with the same pointer identifiers and sort them according to the event time $\systemtimenano$.
Afterwards all events that take place between \TouchDown\ or \TouchPDown\ event action and their respective up event actions with the same pointer ID are accounted as one stroke.
The event \TouchDown\ marks the first touch of the first pointer of a gesture or after a stroke gap.
\TouchPDown\ is fired when further pointers join the set of active pointers while others are already active.
The output of this algorithm is $\touchdata{1}$.

\subsubsection{Motion Data Snuggling}\label{sec:snuggle}
We state that a gesture does start with the first touch and ends with the last touch release before finishing the round by tapping on a button on the screen.
Therefore, we want to cut off the unnecessary motion data before the first touch event and after the last one for each round.
Note that this preserves the motion data for stroke gaps, since they are between strokes, that is only motion data is generated but no touch data.
We call this process \emph{snuggling}, because the motion data gets close to the touch data and encloses it.
Hence, this algorithm computes the cutoff for the motion data sets $\gyrodata{0}$ and $\laccdata{0}$.
Observe that before touching a device and after releasing a pointer, i.e., finger, from the device it moves a bit by the finger motion and the hand holding the device depending on the relative positions.
Therefore we introduce two time offsets as system parameters, $\spmotionstartoffset$ for the offset at the beginning and $\spmotionendoffset$ for the offset at the end of a round.
Previous work \cite{DBLP:conf/mobisys/MiluzzoVBC12} and our tests yield the following values in milliseconds:
\[ \spmotionstartoffset := 150, \qquad \spmotionendoffset := 100. \]
Applying the cutoff yields the data sets $\gyrodata{1}$ and $\laccdata{1}$.

\subsubsection{Motion Fusion Set Computation}\label{sec:computemotionfusion}
Both the gyroscope and accelerometer sensors report data in their own speed and hence are not synchronized, often they even may ``stutter'' a bit.
Even if we tell the sensors to report data every five milliseconds some background processes may keep the CPU busy, such that a positive or negative delay may occur until the data is handed over to \SMAUG.
The task of this algorithm is to merge motion events of $\gyrodata{0}$ and $\laccdata{0}$ that belong together, i.e., have happened at the same time.
To this end, we assume that the workload of the CPU is at a minimum such that in the majority of the cases there is a low delay in the reporting of events that took place at the same point in time.
The algorithm searches for value pairs and aligns these data sets which results into a single synchronous data set of motion events.
Obviously this may result into the situation that some motion events are discarded.
However, this is not a problem due to the huge data amount gathered.

We create a new motion data set $\fusiondata{1}$ containing these fusion values.
An event in this set consists of the following eight values: The gesture identifier, round number $\round$, three values from each motion sensor, and the associated event timestamp.

\subsection{Feature Extraction}\label{sec:featureextraction}
\begin{figure}[!htb]
\centering
\includegraphics[width=0.8\columnwidth]{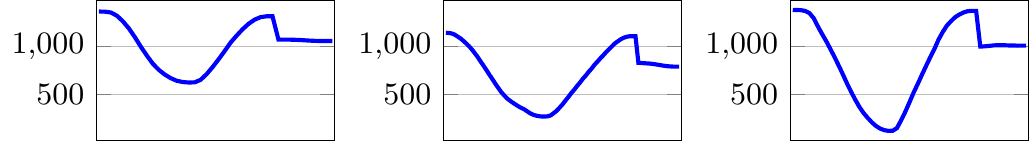}\\
\includegraphics[width=0.8\columnwidth]{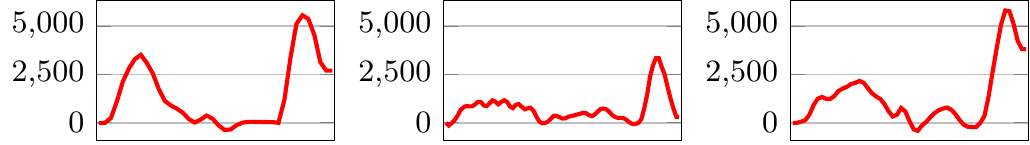}\\[-0.3mm]
\includegraphics[width=0.8\columnwidth]{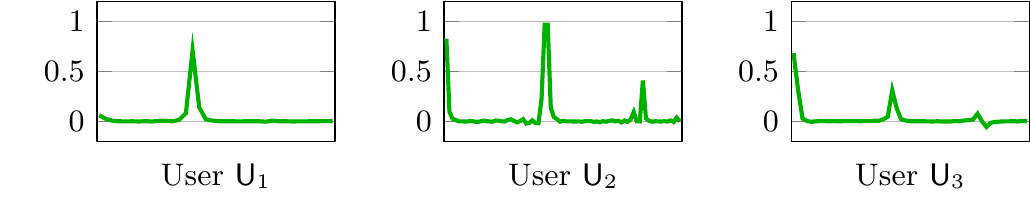}
\caption{Each column represents the inputs of one of three different users ($\mathsf{U}_1$, $\mathsf{U}_2$, or $\mathsf{U}_3$) for the gesture \gesture{Y}.
A graph is generated for each user and feature y-coordinate (top blue graphs), velocity in pixel/second (middle red graphs), and curvature (bottom green graphs).
While the Y-shape is recognizable in the top graph, the other graphs differ more per user, and a combination of all three features allows a user identification with high probability.
To provide distinguishability between more users, or to prevent an impostor, more features will be required for a gesture -- \SMAUG\ employs 320 different features.}
\label{fig:datagraphs}
\end{figure}

After post processing of the data sets, the next step is to extract characteristic features.
A feature can be any value computed from an original data set.
Though it should represent a property of the input, for example the start point coordinates of a stroke or the variance of the pointer size.
For an example of feature comparisons, see Figure~\ref{fig:datagraphs}.
These features can later on be compared to feature values of other data sets.
\SMAUG\ employs not only the features of single sensors, but fuses them in different ways.
The motion sensors are synchronized and interpreted as a single sensor, without excluding individual sensor features.
Data from the touch screen is structured in a way which allows for the extraction of stroke dependent features for each single stroke in addition to features applying to the whole gesture at once.
Merging everything together yields an entangled net of features allowing to distinguish between users.
The main algorithm for this step is denoted by
\[ (\touchdata{1}, \touchdata{2}, \touchdata{3}, \gyrodata{2}, \laccdata{2}, \fusiondata{2})_\round \gets \algFeatureExtraction((\touchdata{1}, \gyrodata{1}, \laccdata{1}, \fusiondata{1})_\round). \]
It applies the following six subsidiary algorithms to the corresponding data in each round $\round$.
As the names indicate the first three algorithms operate on touch related data while the remaining three compute on motion data.

\subsubsection{Touch Event Features}\label{sec:toucheventfeatures}
This algorithm computes within a round $\round$ the features curvature $\curvature\in\R$ and direction $\direction\in\R$ for each stroke $\stroke$ of a touch event.
Additionally, velocity and acceleration in x and y direction are computed as $\velocityX, \velocityY, \accelerationX, \accelerationY \in \R$.

Let $\stroke$ be an arbitrary stroke within the considered round $\round$ and let $I_\stroke$ be the set of indexes that refer to vectors of $\touchdata{1}$ that belong to this stroke.
In the following, let $\feature \in \{x, y, \systemtimenano\} =: \featureset$ denote a feature of $(\touchdata{1})_\round$.
Let $\stroke[\feature] := \bigcup_{i\in I_\stroke}(\touchdata{1}_i[\feature])_\round$ be a short notation for all feature entries of feature $\feature$ in stroke $\stroke$.
Selecting a specific entry $j$ of $\stroke[\feature]$ is denoted by $\stroke_j[\feature]$.

The curvature $\curvature_i$ at point $(\stroke_i[x], \stroke_i[y])$, $1 \leq i \leq |\stroke[x]|$, can be computed as
\begin{align*}
\curvature_i := \frac{4\Phi(y_i)\Theta(\stroke_i[x]) - 4\Phi(\stroke_i[x])\Theta(\stroke_i[y])}{\left(\Theta(\stroke_i[x])^2 + \Theta(\stroke_i[y])^2\right)^{3/2}},
\end{align*}
where $\Phi(\stroke_i[\feature]) := \stroke_{i+1}[\feature]-2\stroke_i[\feature]+\stroke_{i-1}[\feature]$ and 
$\Theta(\stroke_i[\feature]) := \frac{1}{2}\left(\stroke_{i+1}[\feature]-\stroke_{i-1}[\feature]\right)$
with $\feature\in\{x, y\}$.
The direction at point $(\stroke_i[x], \stroke_i[y])$, is defined as
\begin{align*}
\direction_i := \arctan\left(\frac{\stroke_{i+1}[y] - \stroke_i[y]}{\stroke_{i+1}[x] - \stroke_i[x]}\right), \qquad
0 \leq i < |\stroke[x]|.
\end{align*}
Velocity in x- and y-direction is computed by
\begin{align}\label{eq:velocity}
\velocityX_i := \frac{\stroke_i[x] - \stroke_{i-1}[x]}{\stroke_i[\systemtimenano] - \stroke_{i-1}[\systemtimenano]}, \qquad
\velocityY_i := \frac{\stroke_i[y] - \stroke_{i-1}[y]}{\stroke_i[\systemtimenano] - \stroke_{i-1}[\systemtimenano]}, \qquad
1 \leq i \leq |\stroke[x]|.
\end{align}
Acceleration in x- and y-direction of a touch event is defined as follows
\begin{align}\label{eq:acceleration}
\accelerationX_i := \frac{\velocityX_i - \velocityX_{i-1}}{\stroke_i[\systemtimenano] - \stroke_{i-1}[\systemtimenano]}, \qquad
\accelerationY_i := \frac{\velocityY_i - \velocityY_{i-1}}{\stroke_i[\systemtimenano] - \stroke_{i-1}[\systemtimenano]}, \qquad
2 \leq i \leq |\stroke[x]|.
\end{align}
These values are a good representation of the user's behavior as they show how fast and in which direction the user moved her fingers straight or curved.
Other gesture recognition algorithms also employed the same features with reliable results, cf. \cite{DBLP:conf/cns/Sun0ZZ14}.
We store these features as new properties into $\touchdata{1}$.

\subsubsection{Touch Round Features}\label{sec:touchroundfeatures}
Next, we compute global touch features for round $\round$.
These features include
the number of records from the touch screen,
frequency of these records, 
maximum number of pointers touching the screen at the same time,
overall length traveled in pixels,
overall time elapsed in nanoseconds,
number of strokes,
x- and y-coordinates of the center of the gesture box,
and the width and height of the gesture box.
Here, a \emph{gesture box} is the smallest rectangle including the gesture on the touch screen.
With overall length or time we denote the sum of lengths or time of all strokes of this round, respectively.
We store these features in the new data set $\touchdata{2}$ where each row consists of the gesture identifier, round number $\round$, and the feature values mentioned above.
These values have also been used by signature recognition algorithms \cite{DTWBVSBS}.

\subsubsection{Touch Stroke Features}\label{sec:touchstrokefeatures}
This algorithm computes a number of features for each stroke within round $\round$. 
As these computations are done separately for each stroke it is sufficient to explain these for one stroke $\stroke$.
In the following, let
\[ \feature \in \{x, y, \pressure, \size, \curvature, \direction, \velocityX, \velocityY, \accelerationX, \accelerationY\} =: \featureset\]
denote a feature of $(\touchdata{1})_\round$.

First, the minimum and maximum values for each feature sequence are computed as $\min(\stroke[\feature])$ and $\max(\stroke[\feature])$.
Then the following features of the stroke $\stroke$ are computed:
length in nanoseconds, 
length in pixels, 
x- and y-coordinates of the start- and endpoint,
start time and end time,
the sum of pixel length in x and y direction respectively,
width and height of the stroke box, 
and x- and y-coordinates of the center of the stroke box.
As a measurement of central tendency we compute the arithmetic mean $\AM$ and root mean square $\RMS$ as
\begin{align}
\AM\left(\stroke[\feature]\right) &:= \frac{1}{|I_\stroke|}\sum_{i\in I_\stroke} \stroke_i[\feature],\label{eq:AM}\\
\RMS\left(\stroke[\feature]\right) &:= \sqrt{\frac{1}{|I_\stroke|}\sum_{i\in I_\stroke} \left(\stroke_i[\feature]\right)^2}.\label{eq:RMS}
\end{align}
Based on these values we can now compute dispersion metrics such as the variance $\VAR$, standard deviation $\StDev$, and the mean absolute deviation $\MAD$:
\begin{align}
\VAR\left(\stroke[\feature]\right) &:= \frac{1}{|I_\stroke|-1}\sum_{i\in I_\stroke} \left(\stroke_i[\feature] - \AM(\stroke[\feature])\right)^2,\label{eq:VAR}\\
\StDev\left(\stroke[\feature]\right) &:= \sqrt{\VAR\left(\stroke[\feature]\right)},\label{eq:StDev}\\
\MAD\left(\stroke[\feature]\right) &:= \sqrt{\frac{1}{|I_\stroke|-1}\sum_{i\in I_\stroke} \left|\stroke_i[\feature] - \AM(\stroke[\feature])\right|}.\label{eq:MAD}
\end{align}
Next, we compute the third and fourth central moment skewness $\Skew$ and kurtosis $\Kurt$ of each feature sequence of the stroke, respectively.
Skewness describes the asymmetry of a probability distribution and kurtosis is a measurement of peakedness or flatness of a curve.
For example, this can be helpful to distinguish between the gestures of a letter ``U'' and a letter ``V''.
We set
\begin{align}
\Skew\left(\stroke[\feature]\right) &:= \frac{1}{|I_\stroke|-1}\sum_{i\in I_\stroke} \left(\frac{\stroke_i[\feature] - \AM(\stroke[\feature])}{\StDev\left(\stroke[\feature]\right)}\right)^3,\label{eq:Skew}\\
\Kurt\left(\stroke[\feature]\right) &:= \frac{1}{|I_\stroke|-1}\sum_{i\in I_\stroke} \left(\frac{\stroke_i[\feature] - \AM(\stroke[\feature])}{\StDev\left(\stroke[\feature]\right)}\right)^4.\label{eq:Kurt}
\end{align}
In the realm of biometric data processing, this set of features is also used for measuring human behavior \cite{DBLP:journals/comsur/LaraL13}. 
To complete these computations, we finally determine the stroke length in pixels and nanoseconds in percentage of the round length, see Section~\ref{sec:touchroundfeatures}.
We summarize our results in a new data set $\touchdata{3}$ where each row consists of the gesture identifier, the round number $\round$, the stroke number $\stroke$, and each of the computed features.

\subsubsection{Motion Event Features}
The first and second order numerical differentiation are computed by this algorithm for round $\round$ and all sensor values of $\gyrodata{1}, \laccdata{1}, \fusiondata{1}$.
The computations are very similar to equations \eqref{eq:velocity} and \eqref{eq:acceleration}.
These values are joined to the respective data set.

\subsubsection{Motion Round Features}\label{sec:motionroundfeatures}
For each round $\round$ and each motion sensor data set $\gyrodata{1}, \laccdata{1}$, we have nine feature sequences per motion event:
three sensor values,
three values for the first order numerical differentiation,
and three values for the second order numerical differentiation.
To shorten the notation, we represent each sequence by $S[\feature_{i}] := (\motiondata{1}[\feature_{i}])_\round$, where $\feature_{i} \in \{\sensorval_{i}$, $\sensorval_{i}'$, $\sensorval_{i}''\}$.
The number of primes denote the order of the numerical differentiation and $\sensorval_{i}$ denotes the sequence of the $i$-th sensor axis, $1\leq i \leq 3$.
As in the short notation of strokes, by $S_j[\feature_i]$ we denote the $j$-th element of the sequence $S[\feature_i]$.
Note that $\motiondata{1}$ does not include $\fusiondata{1}$ for this algorithm and merely represents either data set $\gyrodata{1}$ or $\laccdata{1}$.

In this algorithm we use the same features as for the touch stroke data algorithm to compute the respective features for each sequence $S[\feature_{i}]$, see equations \eqref{eq:AM}-\eqref{eq:Kurt}.
These features are often used for accelerometer sensor data evaluation \cite{DBLP:journals/comsur/LaraL13}.
We also compute the correlation value between the three axis pairs of each motion sensor.
% as follows:
%\begin{align}
%\Corr_{(g,a)} := \frac{\sum_{k=1}^{|\motiondata{1}|} (\motiondata{(0,2)}_{\feature_i,k} - \AM(\motiondata{(0,2)}[\feature_i]))(\motiondata{(0,2)}_{\feature_j,k} - \AM(\motiondata{(0,2)}_{\feature_j}))}{(|\motiondata{(0,2)}_\round|-1)\cdot\StDev(\motiondata{(0,2)}[\feature_i])\cdot\StDev(\motiondata{(0,2)}_{\feature_j})},
%\end{align}
%where $(g,a)$ denotes the set of all combinations of cross-sensor axis data.
%For example $\Corr_{(1,2)}$ denotes the correlation coefficient of the first sensor value set of the gyroscope data, e.g., yawn, and the second value set of the accelerometer data, e.g., acceleration in x-direction.
%\CG{the Corr formula needs to be changed or removed}
Next, we compute four matrix norm values over $\{\feature_1, \feature_2, \feature_3\}$ for the same differentiation and round $\round$: $1$-norm $||\cdot||_1$ (maximum absolute column sum of the matrix), $\infty$-norm $||\cdot||_\infty$ (maximum absolute row sum of the matrix), Frobenius norm $||\cdot||_\mathrm{F}$ (square root of the squared sum entries of the matrix), and squared $\ell_2$-norm $||\cdot||^2_{\ell_2}$ as follows:
\begin{align*}
||\cdot||_1 &:= \max_{1\leq j \leq 3} \left( \sum_{i=1}^{|S[\feature_{j}]|} |S_i[\feature_{j}]| \right),\\
||\cdot||_\infty &:= \max_{1\leq i \leq |S[\feature_1]|} \left( \sum_{j=1}^3 |S_i[\feature_j]| \right),\\
||\cdot||_\mathrm{F} &:= \sqrt{\sum_{i=1}^{|S[\feature_1]|} \sum_{j=1}^3 S_i[\feature_j]^2},\\
||\cdot||^2_{\ell_2}[i] &:=\sum_{j=1}^3 S_i[\feature_j]^2, \qquad 1 \leq i \leq |S[\feature_1]|.
\end{align*}
Finally, we compute the minimum, maximum, and arithmetic mean from the vector of the squared $\ell_2$-norm $||\cdot||^2_{\ell_2}$.
These values were chosen because they showed good results in classifying user motion on touch screen devices, cf. \cite{DBLP:conf/mobisys/MiluzzoVBC12}.
The frequency of the sampling is not stored on purpose as this depends on external factors like CPU workload which may differ from gesture to gesture.
All resulting feature values are stored in $\gyrodata{2}$ and $\laccdata{2}$, respectively, together with the gesture identifier and the round number $\round$.

\subsubsection{Motion Fusion Features}\label{sec:motionfusionfeatures}
As the two previous algorithms derived features for each sensor data separately the aim of this algorithm is to determine correlations between both gyroscope and accelerometer data sets in $\fusiondata{1}$.
We compute the angle between both motion sensor vectors and its rate of change, also for the first and second order numerical differentiations, and the Pearson correlation coefficients $\PCC$ between all nine pairs of the individual sensor axis.
%\\
%Each axis sequence of a motion sensor in $\motiondata{(1,1)}$ consists of $\motionevents$ values, which are indexed by $i$. For an motion fusion event $i$ we compute the values
%\begin{align*}
%\theta_i :=& \arccos\left(\frac{\sum_{j=1}^3 \motiondata{(1,1)}_{\gyro_j,i} \motiondata{(1,1)}_{\lacc_j,i}}{\sqrt{\sum_{j=1}^3 \left(\motiondata{(1,1)}_{\gyro_j,i}\right)^2\cdot\sum_{j=1}^3 \left(\motiondata{(1,1)}_{\lacc_j,i}\right)^2}}\right),
%\end{align*}
%\begin{align*}
%\phi_i :=& \frac{\theta_i - \theta_{i-1}}{\motiondata{(1,1)}_{\systemtimenano,i} - \motiondata{(1,1)}_{\systemtimenano,i-1}},
%\end{align*}
%where $\theta_i$ is the angle, $\phi_i$ the rate of change of the angle, and $\motiondata{(1,1)}_{\systemtimenano,i}$ the event time in nano seconds.
%These values are now also computed for the first and second order numerical deviations of $g$ and $a$. These resulting six sequences are now appended to the data set $\motiondata{(1,1)}$ yielding $\motiondata{(1,2)}$.
Let $\gyrodataS[i] := (\fusiondata{1}[\text{gyro}_i])_\round$ be the sequence of all gyroscope data of the $i$-th axis in the fusion data set $\fusiondata{1}$.
Furthermore, we define similarly to previous notation $\gyrodataS_k[i]$ as the $k$-th value of the sequence $\gyrodataS[i]$.
We also set $\tilde{\gyrodataS}_k[i] := \gyrodataS_k[i] - \AM\left(\gyrodataS[i]\right)$ and analogously define $\laccdataS[j]$, $\laccdataS_k[j]$, and $\tilde{\laccdataS}_k[j]$ for the accelerometer data.
Note that it holds $|\gyrodataS[i]| = |\laccdataS[j]|$ for any combination of $1\leq i,j\leq 3$ since there is the same amount of sensor data for each event in the fusion data set.
Then, the $\PCC$ coefficients are computed as
\begin{align*}
\PCC_{i,j} := \frac{
\sum_{k=1}^{|\gyrodataS[i]|}
\tilde{\gyrodataS}_k[i]
\tilde{\laccdataS}_k[j]
}{
\left( |\gyrodataS[i]|-1 \right)
\sqrt{\frac{1}{|\gyrodataS[i]|-1} \sum_{k=1}^{|\gyrodataS[i]|} \tilde{\gyrodataS}_k[i]}
\sqrt{\frac{1}{|\gyrodataS[i]|-1} \sum_{k=1}^{|\gyrodataS[i]|} \tilde{\laccdataS}_k[j]}
},
\end{align*}
for $1 \leq i,j \leq 3$, where $\PCC_{ij}$ describes the $\PCC$ values between axis $i$ of the gyroscope and axis $j$ of the accelerometer.
These features have been chosen because they performed well in our tests and have been used before in similar settings \cite{DBLP:conf/mobisys/MiluzzoVBC12}.
The six angle related values are added to $\fusiondata{1}$ while the nine correlation values are added to a new set $\fusiondata{2}$ together with a gesture identifier and round number $\round$.

\subsection{Template Generation}\label{sec:templategeneration}
For the verification process later on, one requires a set of features, called a \emph{template}, that is efficiently comparable for two different gestures.
The template generation algorithm is given as follows
\[ (\touchdataset, \motiondataset)\gets\algTemplateGeneration\left(\metadata, \big\{(\touchdata{1}, \touchdata{2}, \touchdata{3}, \fusiondata{1}, \motiondata{2})_r\big\}_{1\leq \round \leq \sprounds}\right). \]
The template $(\touchdataset, \motiondataset)$ contains all previously computed values required for a gesture comparison and additionally determines limits for the feature values.
Furthermore, meta data are added in order to label the template correctly.
The algorithm consists of six sub-algorithms that are described below. Each of them computes five values: minimum, maximum, standard deviation, median, and arithmetic mean value for each feature sequence.

\subsubsection{Touch DTW Template}\label{sec:touchdtwtemplate}
The algorithm of Section~\ref{sec:determinestrokes} yielded all involved strokes.
Thus, strokes from different rounds can be associated to each other.
This, however, requires that the strokes are always input in the same order.
The goal of this algorithm is to determine for each stroke of the gesture the ``best'' stroke in terms of deviation from the same stroke of the other rounds.
It is based on DTW and can be seen as an extension for multi-touch and multi-stroke recognition of the technique used in \cite{DBLP:conf/chi/LucaHBLH12}.
The input for this algorithm is $\touchdata{1}$.
In a nutshell, it computes for each round and each stroke the distance to the same stroke in the other rounds using DTW.
Therefore, the best stroke is represented by a sequence of the round with the shortest DTW-distance compared to all other rounds. 

For each stroke $\stroke$ and each feature $\feature \in \featureset$ we denote the sequence of its feature values within this stroke as in Section~\ref{sec:toucheventfeatures}, that is by $\stroke[\feature]$.
We denote by $(\stroke[\feature])_\round$ the respective feature values of stroke $\stroke$ in round $\round$.
Next, we compute the distance between feature sequences of all rounds by
\begin{align*}\label{eq:dtwtouch}
\dtw^{\feature}_{\round,\round'} := \DTW\left(\left(\stroke[\feature]\right)_\round, \left(\stroke[\feature]\right)_{\round'}\right)
\end{align*}
for each different pair of rounds $1 \leq \round < \round' \leq \sprounds$.
We also compute the average of the distance for each round $\round$ and feature to all other rounds by $a^{\feature}_\round := \frac{1}{\sprounds} \sum_{j=1}^\sprounds \dtw^{\feature}_{\round, j}$.
Afterwards, the values of $a$ are normalized by dividing them by the maximum $m := \max_{\round=1,\ldots,\sprounds}(a^{\feature}_\round)$ of $a^{\feature}_\round$ over all rounds, resulting in $\overline{a}^{\feature}_\round := a^{\feature}_\round/m$.
Next, the normalized distances are summed up and the result is divided by the number of features which yields
$\overline{\dtw}_\round := \frac{1}{|\featureset|}\sum_{i=1}^{|\featureset|} \overline{a}^{\feature_i}_{\round}, \feature_i\in\featureset$. 
This allows determining the ``best'' stroke instance, i.e., the round where on average the distances of the stroke are the smallest compared to all other rounds, by
$\bestround := \min_{\round=1,\ldots,\sprounds}\left(\overline{\dtw}_\round\right)$.
Using the values computed before we can now determine a kind of ``faultiness'' of the other rounds in respect to the best round $\bestround$ for each feature $\feature\in\featureset$ corresponding to the value $\dtw^\feature_{\bestround, \round}$ for each round $\round\neq\bestround$.
Now we know how ``bad'' a touch sequence of the user may be, depending on the amount of faultiness.
Consequently, for each feature $\feature$ the following five values are computed:
\[\min\left(\left(\dtw^\feature_{\bestround, \round}\right)_{\round=1,\ldots,\sprounds}\right),
\max\left(\left(\dtw^\feature_{\bestround, \round}\right)_{\round=1,\ldots,\sprounds}\right),
\StDev\left(\left(\dtw^\feature_{\bestround, \round}\right)_{\round=1,\ldots,\sprounds}\right),\]
\[\Median\left(\left(\dtw^\feature_{\bestround, \round}\right)_{\round=1,\ldots,\sprounds}\right), \text{and }
\AM\left(\left(\dtw^\feature_{\bestround, \round}\right)_{\round=1,\ldots,\sprounds}\right),\]
where $\Median(I)$ denotes the median of a sequence $I$.
The terms $\StDev$ and $\AM$ are defined in equations~\eqref{eq:StDev} and~\eqref{eq:AM}, respectively.
Observe that these values belong to each stroke instead the whole gesture which allows us to use multi-touch events and multi-strokes.
Finally, we store these values together with a gesture identifier, stroke number $\stroke$, and best round $\bestround$ in a new template data set $\touchdata{4}$.

\subsubsection{Touch Gesture Template}\label{sec:algtouchgesturetemplate}
Let $\featureset_1 := \{$maximum number of pointers at the same time, number of strokes$\}$ and $\featureset_2 := \{$records, overall length in pixels, overall elapsed time in nano seconds, x coordinate of the gesture box center, y coordinate of the gesture box center, box width, box height$\}$ for each round $\round$, see Section~\ref{sec:touchroundfeatures}.
The number of strokes and the maximum number of pointers at the same time are crucial features for defining a gesture.
A deviation in these features results in a completely different structured gesture.
Therefore we call them \emph{strong features} and compute the template value as the rounded down value of $\AM$ of the strong features over all rounds.
Regarding $\featureset_2$, we compute the five template values minimum, maximum, standard deviation, median, and arithmetic mean of $\touchdata{2}[\feature]$ for each $\feature\in\featureset_2$.
We store all computed values in the new template data set $\touchdata{5}$ together with a unique gesture identifier.

\subsubsection{Touch Stroke Template}
For each stroke $\stroke$ and each feature $\feature\in\featureset$ of $\touchdata{3}$, we compute the minimum, maximum, standard deviation, median, and arithmetic mean of the sequences $\touchdata{3}[\feature]$.
The results are stored together with a unique gesture identifier and stroke $\stroke$ in the template data set $\touchdata{6}$.

\subsubsection{Motion Round Template}
For each of the two motion sensors, this algorithm takes all features described in Section~\ref{sec:motionroundfeatures} of $\gyrodata{2}$ and $\laccdata{2}$, and computes for each feature sequence its minimum, maximum, standard deviation, median, and arithmetic mean.
Finally, these values are stored together with a unique gesture identifier in the respective template data sets $\gyrodata{3}$ and $\laccdata{3}$.

\subsubsection{Motion Fusion Template}
This algorithm operates on the fusion data set $\fusiondata{2}$ and the features computed in Section~\ref{sec:motionfusionfeatures}.
For each feature sequence, the minimum, maximum, standard deviation, median, and arithmetic mean are computed. 
The resulting values are stored together with a unique gesture identifier in the template data set $\fusiondata{3}$.

\subsubsection{Motion DTW Template}\label{sec:motiondtwtemplate}
This algorithm works analogue to Section~\ref{sec:touchdtwtemplate}.
However, we now employ the fusion data set $\fusiondata{1}$ instead of touch data.
We again get the best round $\bestround$ for all distances measured for each feature.
Finally, we compute the minimum, maximum, standard deviation, median, and arithmetic mean values for each feature distance sequence.
Afterwards, we store these values together with a unique gesture identifier and best round $\bestround$ in the template data set $\fusiondata{4}$.

\bigskip
Finally, the algorithm $\algTemplateGeneration$ returns the template $(\touchdataset, \motiondataset)$, where
\begin{align*}
\touchdataset &:= \cup_{\round=1}^\sprounds \cup_{i=0}^3 \left( (\touchdata{i})_\round\right) \cup \left(\cup_{i=4}^6 \touchdata{i}\right),\\
\motiondataset &:= \cup_{\round=1}^\sprounds \left( \cup_{i=0}^2 \left( (\gyrodata{i})_\round \cup (\laccdata{i})_\round \right)  \cup \left( \cup_{i=1}^2 (\fusiondata{i})_\round \right) \right) \cup \gyrodata{3} \cup \laccdata{3} \cup \left( \cup_{i=3}^4 \fusiondata{i} \right).
\end{align*}
Please note that we implicitly also store a gesture identifier derived from meta data $\metadata$ in the template.

\subsection{Computing Individualized Weights}\label{sec:computeweights}
Given the template as described above, the verification process could work as follows: given a new input from some user $\unknownuser$, just check if it is sufficiently similar to the templates.
Not surprising, some features will be more characteristic than others.
However, our experiments showed that the expressiveness of features also depends on the user and the entered gesture.
In other words, each feature has a different relevance for each user, gesture, and stroke.
For example, for a user $\user_1$ the x- and y-coordinates could be more important than the velocity of drawing the gesture since the user's speed of performing the input varies while he is quite accurate in reproducing the gesture.
But for another user $\user_2$, it may just be the other way around.

This makes it necessary to introduce and assign to each feature its  \emph{feature weight}, expressing the level of relevance for the authentication process.
Therefore, the feature weights cannot be some global system parameters.
In fact, \SMAUG\ identifies dynamically these weights during the enrollment phase depending on the stability of the input.
On top of these individual weights we also introduce \emph{feature tiers}.
Our experiments showed that across all users, certain groups of features always perform slightly better for honest user recognition than other features.
However, latter are getting useful when preventing strong adversaries as allowed by our adversary model, see Section~\ref{sec:adversarymodel}.
A feature tier represents a second weighting for a class of similar features and each feature is assigned exactly to one tier.
Based on our experiments, we decided to have three tiers in total.
In comparison, feature weights are computed individually, and feature tier weights are pre-computed values based on our experiments.
Summing up, the weight computation algorithm is denoted by
\[ (\weightsset, \faultindicator, \weightindicator) \gets \algComputeWeights\left(\touchdataset, \motiondataset\right) \]
which takes the gesture template as input and outputs a set of weights $\weightsset$ and also two indicators $\faultindicator$, $\weightindicator$.
It dynamically computes the feature weights for a gesture and each of its strokes while taking the global feature tier values into account.
The algorithm can be described briefly as follows: by cross-validation we compare the gestures template to each gesture round input performed by the user and store the deviations in a novel \emph{fault container} $\faultcontainer$.
Afterwards, we compute the feature weights based on the information in the error container $\faultcontainer$.
The feature weight computation depends on the number of overall rounds during the enrollment phase $\sprounds$, the more data the better the weight estimation.

\subsubsection{Internal Cross-Validation}\label{sec:PerformChecks}
The cross-validation step performs comparison checks for each of the original user inputs $\{(\touchdata{0}, \gyrodata{0}, \laccdata{0})_\round\}_{1\leq \round \leq \sprounds}$ to the gesture template $(\touchdataset, \motiondataset)$ and outputs the fault container $\faultcontainer$.

There are three possible types of comparison which we denote by $\comparisontypes:=\{\LB, \UB, \EQ\}$, where $\LB$ stands for ``lower bound'', $\UB$ for ``upper bound'', and $\EQ$ for ``equality'', respectively.
The comparisons are defined as follows:
\begin{align}
\LB(v, a, m, s) &:= 
\begin{cases}
0, & v < \frac{1}{2} \left(a + m \right) - s,\\
1, & \text{else},
\end{cases}\label{eq:cmplb}
\\
\UB(v, a, x, s) &:= 
\begin{cases}
0, & v > \frac{1}{2} \left(a + x \right) + s,\\
1, & \text{else},
\end{cases}\label{eq:cmpub}
\\
\EQ(v, w) &:= 
\begin{cases}
0, & v=w,\\
1, & \text{else}.
\end{cases}\label{eq:cmpeq}
\end{align}
Here, $v$ and $w$ refer to a feature value, and $a$ to the arithmetic mean, $m$ to the minimum, $x$ to the maximum, and $s$ to the standard deviation stored in the according template.
If a comparison fails, the output will be $1$, and otherwise $0$.
The form of these comparisons are a result of our extensive testing phase.
This gives a very good estimation of the deviation in the ``environment'' of features caused by a user's behavior.
The kind of comparison performed depends on the feature, but the result is always binary: either a comparison is passed or otherwise failed.
Features failing a comparison get stored in $\faultcontainer$ together with their comparison type $\comparison$.
The cross validation algorithm consists of the following seven subsidiary algorithms, which are performed for each $\round=1,\ldots,\sprounds$.

\paragraph{Strong Features Check}\label{sec:algcheckstrongfeatures}
This algorithm compares the strong features of the gesture, that is
``number of strokes'' and ``maximum number of pointers at the same time''.
Both features are denoted by the set $\featureset_1$, see Section~\ref{sec:algtouchgesturetemplate}.
The comparison is done as an equality check between data set $\touchdata{2}$ (see Section~\ref{sec:touchroundfeatures}) and template data set $\touchdata{5}$ (see Section~\ref{sec:algtouchgesturetemplate}).
Please observe, that $\touchdata{2}[\feature]$ is a single value.
If a feature is not the same in both sets, i.e., the test
\begin{align*}
\EQ\left(
\left(\touchdata{2}[\feature]\right)_\round,\ 
\touchdata{5}[\feature] \right),
\quad \feature\in\featureset_1,
\end{align*}
as defined in equation~\eqref{eq:cmpeq} fails and equals $1$, the feature $\feature$ and round $\round$ together with comparison type $\EQ$ get stored in $\faultcontainer$.

\paragraph{Touch Gesture Check}\label{sec:checktouchgesture}
This algorithm checks if the corresponding features of the set $(\touchdata{2})_\round$ do not ``deviate too much'' from the template data set.
That is, each feature $\feature\in\featureset_2$ is verified according to a lower and upper bound, where $\featureset_2$ is defined in Section~\ref{sec:algtouchgesturetemplate}.
Hence, the comparisons from equation~\eqref{eq:cmplb} and equation~\eqref{eq:cmpub} are parametrized as
\begin{align*}
&\LB\left(
\left(\touchdata{2}[\feature]\right)_\round,\ 
\touchdata{5}[\AM_\feature],\ 
\touchdata{5}[{\min}_\feature],\ 
\touchdata{5}[\StDev_\feature]
\right),\\
&\UB\left(
\left(\touchdata{2}[\feature]\right)_\round,\ 
\touchdata{5}[\AM_\feature],\ 
\touchdata{5}[{\max}_\feature],\ 
\touchdata{5}[\StDev_\feature]
\right),
\quad \feature\in\featureset_2.
\end{align*}
The notation $\touchdata{5}[\AM_\feature]$ refers to the arithmetic mean of the feature $\feature$ stored in the data set $\touchdata{5}$, and analog minimum, maximum, and standard deviation.
If a test fails, i.e., $\LB=0$ or $\UB=0$, the feature $\feature$, round $\round$, and bound ($\LB$ or $\UB$) are added to the fault container $\faultcontainer$.

\paragraph{Touch Stroke Check}
For each stroke $\stroke$ in $\touchdata{6}$ we compare the feature values $\feature\in\featureset$ with the ones from the value of the set $(\touchdata{3})_\round$.
This is done by checking the lower and upper bounds using equations (\ref{eq:cmplb}) and (\ref{eq:cmpub}), respectively.
The single feature value is given by $(\touchdata{3}[\feature])_\round$ while the according values of $\AM$, $\min$, $\max$, and $\StDev$ are provided directly by $\touchdata{6}[\feature]$, that is
\begin{align*}
&\LB\left(
\left(\touchdata{3}[\feature]\right)_\round,\ 
\touchdata{6}[\AM_\feature],\ 
\touchdata{6}[{\min}_\feature],\ 
\touchdata{6}[\StDev_\feature]
\right),\\
&\UB\left(
\left(\touchdata{3}[\feature]\right)_\round,\ 
\touchdata{6}[\AM_\feature],\ 
\touchdata{6}[{\max}_\feature],\ 
\touchdata{6}[\StDev_\feature]
\right),
\quad \feature\in\featureset.
\end{align*}
On failing a test, the respective feature $\feature$, round $\round$, comparison type, and stroke $\stroke$ are added to the fault container $\faultcontainer$.

\paragraph{Touch Stroke DTW Check}\label{sec:checktouchstrokedtw}
This algorithm takes the set $(\touchdata{1})_\bestround$ as an input, where $\bestround$ denotes the best round, see Section~\ref{sec:touchdtwtemplate}.
The comparison data set is $(\touchdata{1})_\round$.
The following two steps are done for each stroke $\stroke$ of the gesture.

First, we check for two features whether they are equal, namely $\featureset_3 := \{$``pointer identifier'', ``pointer numbers''$\}$.
The values of $\featureset_3$ are also treated as strong features.
Hence, if they are not equal in $(\touchdata{1})_\bestround$ and $(\touchdata{1})_\round$ for the given stroke $\stroke$, they get added to the set $\faultcontainer$ together with $\EQ$.
Second, the DTW distances of feature sequences are considered.
Now the ``best'' round of the gesture inputs gets compared to all other data sets, i.e., $(\touchdata{1}[\feature])_\bestround$ and the data set $(\touchdata{1}[\feature])_\round$ which yields the DTW distance.
Then, this is compared to the values of $\touchdata{4}$ as
\begin{align*}
\UB\left(
\dtw^\feature_{\bestround, \round},\ 
\touchdata{4}[\AM_\feature],\ 
\touchdata{4}[{\max}_\feature],\ 
\touchdata{4}[\StDev_\feature]
\right),
\quad \feature\in\featureset.
\end{align*}
If the distance is not below the upper bound given by equation \eqref{eq:cmpub}, it is considered as an error and the feature is inserted into $\faultcontainer$ together with comparison type $\UB$, round $\round$, and stroke $\stroke$.
Note that we do not have to check for a lower bound as it holds for at least one test data that one feature perfectly matches, i.e., yielding a distance of $0$.

\paragraph{Motion Gesture Check}
The motion gesture check uses a set of features which we denote by $\featureset_\text{MG}$.
A brief description of its 90 features can be found in Section~\ref{sec:motionroundfeatures}.
For each the gyroscope and the accelerometer data, this algorithm will check the features of the set $\featureset_\text{MG}$ in comparison between the values of $\gyrodata{2}$ and $\gyrodata{3}$, as well as $\laccdata{2}$ and $\laccdata{3}$.
Seven features are computed for each axis value and its first and second numerical differentiation.
Also the correlation between each axis pair as well as four matrix norm values of the axis values and differentiations are part of this set.
Then, it performs for each motion sensor and each feature the checks given by equations \eqref{eq:cmplb} and \eqref{eq:cmpub} as follows:
\begin{align*}
&\LB\left(
\left(\gyrodata{2}[\feature]\right)_\round,\ 
\gyrodata{3}[\AM_\feature],\ 
\gyrodata{3}[{\min}_\feature],\ 
\gyrodata{3}[\StDev_\feature]
\right),\\
&\UB\left(
\left(\gyrodata{2}[\feature]\right)_\round,\ 
\gyrodata{3}[\AM_\feature],\ 
\gyrodata{3}[{\max}_\feature],\ 
\gyrodata{3}[\StDev_\feature]
\right),\\
&\LB\left(
\left(\laccdata{2}[\feature]\right)_\round,\ 
\laccdata{3}[\AM_\feature],\ 
\laccdata{3}[{\min}_\feature],\ 
\laccdata{3}[\StDev_\feature]
\right),\\
&\UB\left(
\left(\laccdata{2}[\feature]\right)_\round,\ 
\laccdata{3}[\AM_\feature],\ 
\laccdata{3}[{\max}_\feature],\ 
\laccdata{3}[\StDev_\feature]
\right),
\quad \feature\in\featureset_\text{MG}.
\end{align*}
The feature $\feature$, round $\round$, and comparison type are added to $\faultcontainer$ whenever a check fails.

\paragraph{Motion Fusion Check}
For this check we are comparing fusion data to the fusion template data.
The feature set is given by $\featureset_\text{MF}$ and includes correlation between all nine axis pairs of gyroscope and accelerometer.
Furthermore, the angle between both motion vectors, its rate of change, velocity, velocity rate of change, acceleration, and acceleration rate of change complete this feature set.
Using equations \eqref{eq:cmplb} and \eqref{eq:cmpub}, we again check the lower and upper bounds by comparing the original value to a combination of template values as
\begin{align*}
&\LB\left(
\left(\fusiondata{2}[\feature]\right)_\round,\ 
\fusiondata{3}[\AM_\feature],\ 
\fusiondata{3}[{\min}_\feature],\ 
\fusiondata{3}[\StDev_\feature]
\right),\\
&\UB\left(
\left(\fusiondata{2}[\feature]\right)_\round,\ 
\fusiondata{3}[\AM_\feature],\ 
\fusiondata{3}[{\max}_\feature],\ 
\fusiondata{3}[\StDev_\feature]
\right),
\quad \feature\in\featureset_\text{MF}.
\end{align*}
Feature $\feature$, round $\round$, and comparison type are added to the fault container $\faultcontainer$ if a test fails.

\paragraph{Motion Fusion DTW Check}
In this algorithm we employ the feature set $\featureset_\text{MD}$ given by the angle between both motion vectors, its rate of change, velocity, velocity rate of change, acceleration, and acceleration rate of change.
Next, we perform the upper bound check over $(\fusiondata{1})_\round$, $(\fusiondata{1})_\bestround$, and $\fusiondata{4}$ as
\begin{align*}
\UB\left(
\dtw^\feature_{\bestround, \round},\ 
\fusiondata{4}[\AM_\feature],\ 
\fusiondata{4}[{\max}_\feature],\ 
\fusiondata{4}[\StDev_\feature]
\right),
\quad \feature\in\featureset_\text{MD}.
\end{align*}
The feature $\feature$, round $\round$, and comparison type $\UB$ are added to the fault container $\faultcontainer$ if a test fails.

\subsubsection{Weight Calculation}\label{sec:algweightcalculation}
This algorithm takes the fault container $\faultcontainer$ as input and outputs the weight set $\weightsset$ for all features considered, that is for both the gesture as a whole and also each stroke.
Recall that the subsidiary algorithms of the internal cross-validation algorithm performed one or two checks, depending on the feature, see Section~\ref{sec:PerformChecks}.
Therefore, for some features up to two faults may occur, namely failing on both comparisons $\LB$ and $\UB$.
In contrast, tests for DTW related features and strong features perform only a single comparison, namely $\UB$ or $\EQ$, respectively.
We call a pair in $\faultcontainer$ consisting of a feature $\feature$ and an associated comparison type $\comparison\in\comparisontypes$ \emph{comparison feature} $\compfeature := (\feature, \comparison)\in\faultcontainer$.
These are the values which have been added in the previously performed internal cross-validation algorithm.

Now, we are going to explain how for each $\compfeature$ a corresponding weight is computed.
In principle, the weight is defined as the frequency that $\compfeature$ did not fail, multiplied by a constant that depends on the tier of $\compfeature$.
More precisely, let $\compfeaturesetg$ denote all comparison features for the global gesture independent of strokes, and $\compfeaturesets$ all comparison features exclusively for a single stroke. 
Furthermore, let $\strokes$ denote the number of all strokes in this gesture.
For a fault container $\faultcontainer$, a comparison feature $\compfeature\in\compfeaturesetg$ and a round $\round\in[\sprounds]$, the term $\faultcontainerg(\compfeature,\round)$ is equal to $1$ if the corresponding fault occurred in $\faultcontainer$ with respect to round $\round$, and is equal to $0$ otherwise.
This is according to the outputs of the three comparisons which output $1$ if a comparison fails, see equations \eqref{eq:cmplb}, \eqref{eq:cmpub}, and \eqref{eq:cmpeq}.
Moreover, we define by $\faultcontainerg(\compfeature):=\sum_{\round\in[\sprounds]} \faultcontainerg(\compfeature,\round)$ the number of occurrences of $\compfeature$ in $\faultcontainer$ for $\compfeature\in\compfeaturesetg$.
In other words, $\faultcontainerg(\compfeature)$ yields how often the feature $\feature$ did not pass a comparison check during the internal cross-validation phase.
Observe that if the fault container $\faultcontainer$ contains only one round, e.g., as it is the case for the verification phase later on, it holds $\faultcontainerg(\compfeature,\round)=\faultcontainerg(\compfeature)$.  
Likewise, we define $\faultcontainers(\compfeature,\round)$ for a comparison feature $\compfeature\in\compfeaturesets$ and a stroke $\stroke\in[\strokes]$, and set $\faultcontainers(\compfeature) := \sum_{\round\in[\sprounds]} \faultcontainers(\compfeature,\round)$.
 
As explained in Section~\ref{sec:computeweights}, all features are partitioned into tiers, where one feature belongs to exactly one tier.
A tier is a weight control for the features and their respective weights.
We consider three different tiers $\tierset1$, $\tierset2$, and $\tierset3$.
By $\tierset_\feature$ we denote the tier $\tierset$ of feature $\feature$, which represents one of the system values $\sptierone$, $\sptiertwo$, or $\sptierthree$. 
We determined in tests which features should be assigned to which tier.
Because of space restrictions, we cannot list the partitioning here.
In general, the more directly the features are derived from user behavior, the higher the tier.
We set the following system parameters, resulting from our experiments:
\[ \sptierone := 0.5, \qquad \sptiertwo := 2, \qquad \sptierthree := 4. \]
Then, for the comparison features, their weights are defined as follows: 
\begin{align}
\weight^{\compfeature}_0 &:= \tierset_\feature \left(1 - \frac{\faultcontainerg(\compfeature)}{\sprounds}\right), \quad \compfeature\in \compfeaturesetg,\label{eq:weightg}\\
\weight^{\compfeature}_\stroke &:= \tierset_{\feature}\left(1 - \frac{\faultcontainers(\compfeature)}{\sprounds}\right), \quad \compfeature\in \compfeaturesets, \quad \stroke\in[\strokes].\label{eq:weights}
\end{align}
We define by $\faultindicator[\round]$ a \emph{fault sum indicator} for round $\round$.
This is the sum of all comparison faults in fault container $\faultcontainer$ which occurred during round $\round$.
Similarly, a \emph{weight sum indicator} $\weightindicator[\round]$ contains the sum of all weights for all faults of round $\round$.
More precisely, this is computed as follows:
\begin{align*}
\faultindicator[\round] &:=
\sum_{\compfeature\in\compfeaturesetg} \weight^{\compfeature}_0 + \sum_{\stroke=1}^{\strokes} \sum_{\compfeature\in\compfeaturesets} \weight^{\compfeature}_\stroke\\
\weightindicator[\round] &:=
\sum_{\compfeature\in\compfeaturesetg} \faultcontainerg(\compfeature, \round) \cdot \weight^{\compfeature}_0 +
\sum_{\stroke=1}^{\strokes} \sum_{\compfeature\in\compfeaturesets} \faultcontainers(\compfeature, \round) \cdot \weight^{\compfeature}_\stroke
\end{align*}
Next, we compute the arithmetic mean and deviation of both indicators over all rounds according to equation \eqref{eq:AM} and equation \eqref{eq:StDev}, respectively.
This yields the values $\AM(\faultindicator)$, $\AM(\weightindicator)$, $\StDev(\faultindicator)$, and $\StDev(\weightindicator)$.
Finally, we compute the global fault sum indicator $\faultindicator$ and weight sum indicator $\weightindicator$ as
\begin{align}
\faultindicator &:= \AM(\faultindicator) + \StDev(\faultindicator),\label{eq:faultindicator}\\
\weightindicator &:= \AM(\weightindicator) + \StDev(\weightindicator).\label{eq:weightindicator}
\end{align}
In other words, these formulas yield the mean value of overall errors occurred and also their according weight sums.
The ultimate goal for these values is to give a numeric representation of the quality of the user inputs which will make it comparable to new inputs, i.e., authentication attempts.
We have chosen the combination of arithmetic mean and standard deviation as the best performing one.
For example, using only the arithmetic mean is not robust enough, since ``small'' outliers are excluded, but our extensive experiment phase showed that they are definitely part of a legitimate user.
On the other hand, if we would only use the maximum here, even the worst of all inputs of a user is still valid.
Similarly, a similar equation as in the comparison bounds, i.e., employing minimum or maximum additionally, is still too forgiving for the later verification phase.

We store $\weightindicator$, $\faultindicator$, and all weights $\weight$ of the comparison features (see equations~\eqref{eq:weightg} and~\eqref{eq:weights}) in the set denoted by $\weightsset$.

\subsection{Thresholds}\label{sec:threshold}
Thresholds are crucial to distinct between a valid authentication try and an impostor later on.
We now describe the algorithm
\[  (\threshold_1,\threshold_2) \gets \algThreshold(\weightsset). \]
Threshold $\threshold_1$ is required for the weight sum of feature faults occurred and $\threshold_2$ is a threshold for the number of feature faults occurred. 
Their values depend on the number of strokes $\strokes$ of the gesture as well as the weight sum indicator $\weightindicator$ and fault sum indicator $\faultindicator$, computed in equation~\eqref{eq:weightindicator} and~\eqref{eq:faultindicator}, respectively.
Both values has been stored in the set $\weightsset$.
Note that the more strokes a gesture has, the more faults may occur. 
Therefore, the number of strokes $\strokes$ directly influences the number of faults and at the same time their weights.
The thresholds are computed as follows:
\begin{align}
\threshold_1 &:= \weightindicator\spfaultweightmult + \spfaultweightadd(1 + \strokes),\label{eq:threshold1}\\
\threshold_2 &:= \faultindicator\spfaultnumbermult + \spfaultnumberadd(1 + \strokes).\label{eq:threshold2}
\end{align}
We introduce the four security parameters of \SMAUG\ here, which allow us to cope with the different modes, see Section~\ref{sec:modes}.
These are $\spfaultweightmult$ and $\spfaultnumbermult$ as multiplicative values for each threshold, while $\spfaultweightadd$ and $\spfaultnumberadd$ are additive values for each threshold.
If a background image is used, the security parameters should be different than without picture, because also the attacker can remember the gesture more easily.
Without a picture the user has no orientation where exactly to hit the screen, therefore the values should lead to more forgiveness.
Furthermore, the thresholds also depend on the structure of the gesture, i.e., if it was single- or multi-touch.
This is due to the fact that single-touch is easier to replicate.
In our tests we computed the values shown in Table~\ref{tbl:secparams}.
\begin{table}[htb]
\centering
\begin{tabular}{c|cc}
\toprule
\textbf{Background Image} & \textbf{Single-Touch} & \textbf{Multi-Touch}\\
\midrule
\multirow{2}{*}{Yes} & $\spfaultweightmult = 2.5$, $\spfaultweightadd = 6$, & $\spfaultweightmult = 2.2$, $\spfaultweightadd = 4$,\\
 & $\spfaultnumbermult = 1.9$, $\spfaultnumberadd = 6$ & $\spfaultnumbermult = 2.1$, $\spfaultnumberadd = 4$\\
\midrule
\multirow{2}{*}{No} & $\spfaultweightmult = 3$, $\spfaultweightadd = 7$, & $\spfaultweightmult = 2.5$, $\spfaultweightadd = 8$,\\
 & $\spfaultnumbermult = 2.1$, $\spfaultnumberadd = 7$ & $\spfaultnumbermult = 2.5$, $\spfaultnumberadd = 8$\\
\bottomrule
\end{tabular}
\vspace*{2mm}
\caption{Default values of \SMAUG's security parameters, which depend on background image and maximum number of active pointers at the same time.}
\label{tbl:secparams}
\end{table}

\section{Verification Phase}\label{sec:verificationphase}
In this section we will explain the verification phase in detail.
We will stick to the algorithmic description of Algorithm~\ref{alg:verification} and explain the five subsidiary algorithms.
In more detail, for each authentication try $\authtry$, $0\leq\authtry\leq\sprepeat$, the while loop is executed until a valid input was made.
As soon as the verification returns $\True$, the algorithm outputs the decision $\decision$ with value $\True$.
If the authenticating user fails in all $\sprepeat+1$ trials, the authentication fails completely.
Then, a fallback mechanism will handle the next steps.
For a brief discussion on this see Section~\ref{sec:instantiationnote}.
However, the value of $\sprepeat$ should be chosen relatively small in order to prevent repeated testing, but high enough for a reasonable usability, such that a mistake by an honest user is not immediately punished.
We think that $\sprepeat := 3$ is a good value and set this as default.
To summarize this, the goal of the verification phase is to determine if an authentication attempt was successful.

\subsection{Data Gathering \& Post Processing}\label{sec:testdgpp}
For an authentication attempt $\authtry$, $1\leq\authtry\leq\sprepeat$, the possibly unknown user $\unknownuser$ enters a gesture $\anyunknowngesture$ which results in the \emph{test} data sets $(\testmetadata, \testtouchdata{0}, \testgyrodata{0}, \testlaccdata{0})$.
This is done exactly the same way as described in Section~\ref{sec:datagathering}.
Then, post processing is applied, which is in fact the same as in Section~\ref{sec:postprocessing}.
This yields the test data sets $\testtouchdata{1}$ and $\testmotiondata{1}$.

\subsection{Test Feature Extraction}\label{sec:testfe}
This step consists of extracting features from $\testtouchdata{1}$ and $\testmotiondata{1}$.
Here, the previously presented algorithm $\algFeatureExtraction$ from Section~\ref{sec:featureextraction} can be employed analogous.
However, please note that in this case we only have one data set instead of $\sprounds$ sets, since only one input round in given.
In other words, for an authentication attempt the gesture is performed \emph{once} instead of $\sprounds$ times as in the enrollment phase.
Therefore we slightly alternate the algorithm and reduce all feature computations to one single round.
We also omit the output of $\testtouchdata{1}$.
Next, we denote the final algorithm for test feature extraction by
\[ (\testtouchdata{2}, \testmotiondata{2}) \gets \algTestFeatureExtraction(\testtouchdata{1}, \testmotiondata{1}). \]
This already yields all values required for a comparison of the test data set features and the gesture template data sets computed in the enrollment phase.

\subsection{Validation}\label{sec:validationXXX}
By employing a single cross-validation, the validation algorithm computes the number of all faults, being the features which do not fulfill requested bound checks, and the weights of those faults.
We compute
\[ (\testweightindicator, \testfaultindicator) \gets \algValidation((\testtouchdata{2}, \testmotiondata{2}), (\touchdataset, \motiondataset), \weightsset),\]
being the same algorithm as given in Section~\ref{sec:PerformChecks}.
In this case, the algorithm yields a fault container $\testfaultcontainer$.
Then, the corresponding test fault sum indicator $\testweightindicator$ and test fault number indicator $\testfaultindicator$ are computed.
This is accomplished by the formulas \eqref{eq:weightindicator} and \eqref{eq:faultindicator}, respectively, but with the difference that the test fault container $\testfaultcontainer$ is used instead of $\faultcontainer$.
It is important to note that the computation of $\testweightindicator$ uses the weights stored in $\weightsset$ (see Section~\ref{sec:algweightcalculation}).
This ensures that the gesture $\anyunknowngesture$ is interpreted as an input of the legitimate user $\user$.
If it was the user $\user$, it will fit very well.
But if an impostor tried to login as $\user$, the weights will mess up the indicator computation in the sense that the value is far off from the one by the legitimate user $\user$.

\subsection{Verification}\label{sec:verificationXXX}
As a final step of the verification phase, we execute the algorithm
\[ \decision \gets \algVerification(\threshold_1, \threshold_2, \testweightindicator, \testfaultindicator). \]
The thresholds $\threshold_1, \threshold_2$ have been computed in equations~\eqref{eq:threshold1} and \eqref{eq:threshold2}, and both indicators $\testweightindicator, \testfaultindicator$ in the previous section.
We set the initial value of the binary decision variable $\decision$ to $\False$.
The comparison then is the test of the following logical equation:
\begin{align*}
\decision = \left(\faultsweight \stackrel{?}{\leq} \threshold_1 \wedge \faultscount \stackrel{?}{\leq} \threshold_2\right).
\end{align*}
The output of $\decision$ is changed to $\True$ if the equation holds.
This concludes the verification phase and determines if the authentication attempt was successful, i.e., $\unknownuser = \user$ and $\anyunknowngesture = \anygesture$ for any user and gesture.

\section{Experiments}\label{sec:experimentsXXX}
In this section we first describe our implementation and test setup for the participants.
Afterwards we give a discussion on the experiment results.

\subsection{Implementation}\label{sec:implementation}
We implemented \SMAUG\ on a Nexus 5 device running on Android 5.1.1.
Users are registered in this app by unique identifiers.
A user can create as many gestures as desired and manage them in the app.
To simulate an authentication process, a user is able to choose any user known to the system and any of her gestures during the login phase.
Then she can try to log in and receives a positive (login successful) or negative (login failed) reply as feedback.
The data gathered by the users were stored in text files and evaluated on a laptop (1.8GHz, SSD), to make data handling for various testing processes easier.
On this computer we used a database to handle the data and executed all algorithms except for the data gathering step, which was always running on the smartphone.
This especially allows for bulk evaluations of already performed authentication attempts, i.e., ``old'' input data already transferred onto the computer, even if alterations on the algorithm were being made during development and the testing phase.
A single gesture needs about 1MB of space and a single authentication is evaluated in less than 0.1 seconds, depending slightly on the size of the gesture.
%The source code of the implementation can be found on Github at http://github.com/CAR/smaug/

\subsection{Setup}\label{sec:testsetup}
\begin{figure}[ht]
\begin{center}
\subfigure[Input of gesture \gesture{A} with helpful background image.]{
\includegraphics[width=0.29\textwidth]{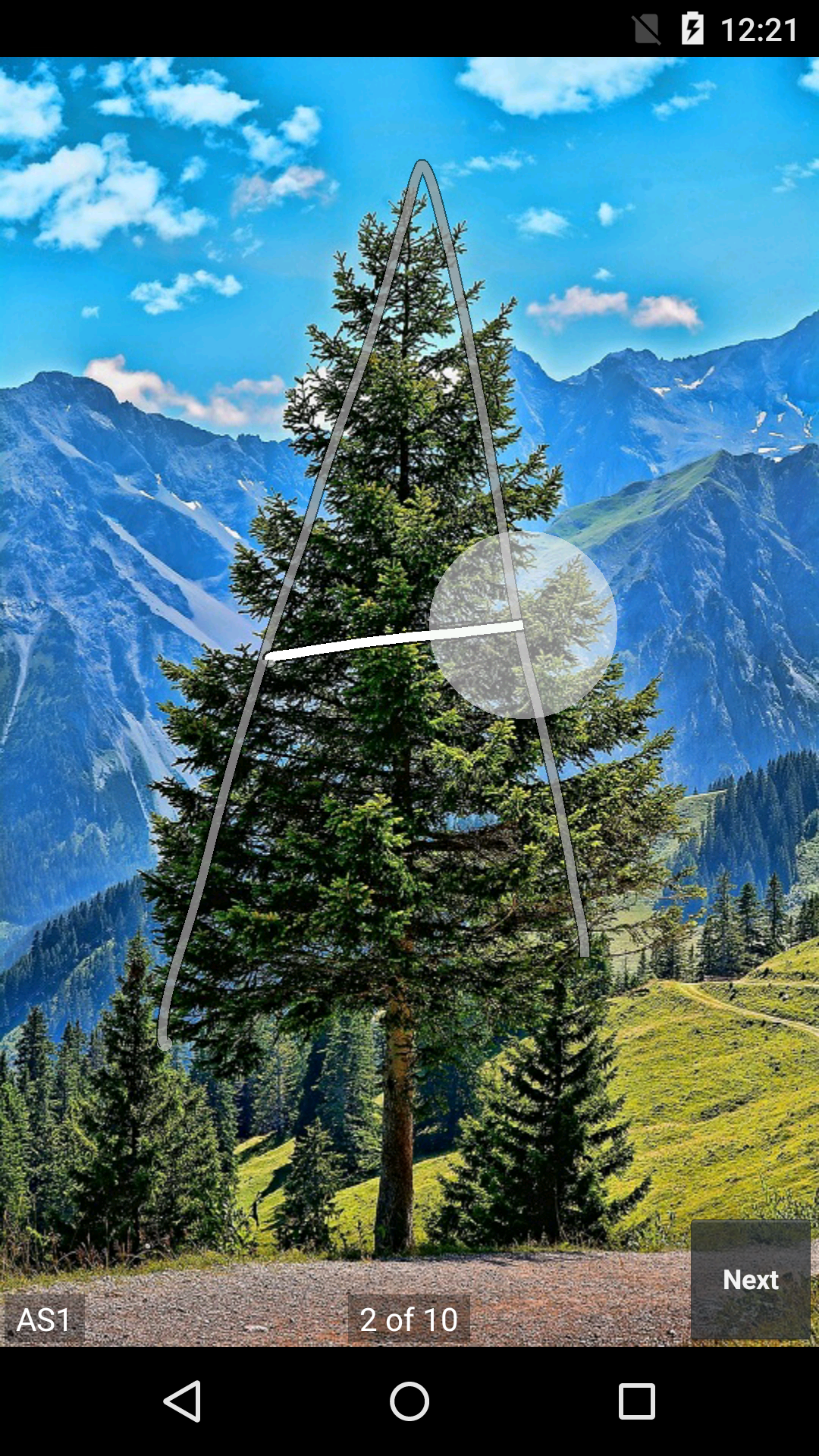}
\label{fig:inputgesturea}
}
\hfill
\subfigure[Support for multi-stroke and multi-touch gestures is support as shown on the input of gesture \gesture{DC}.]{
\includegraphics[width=0.29\textwidth]{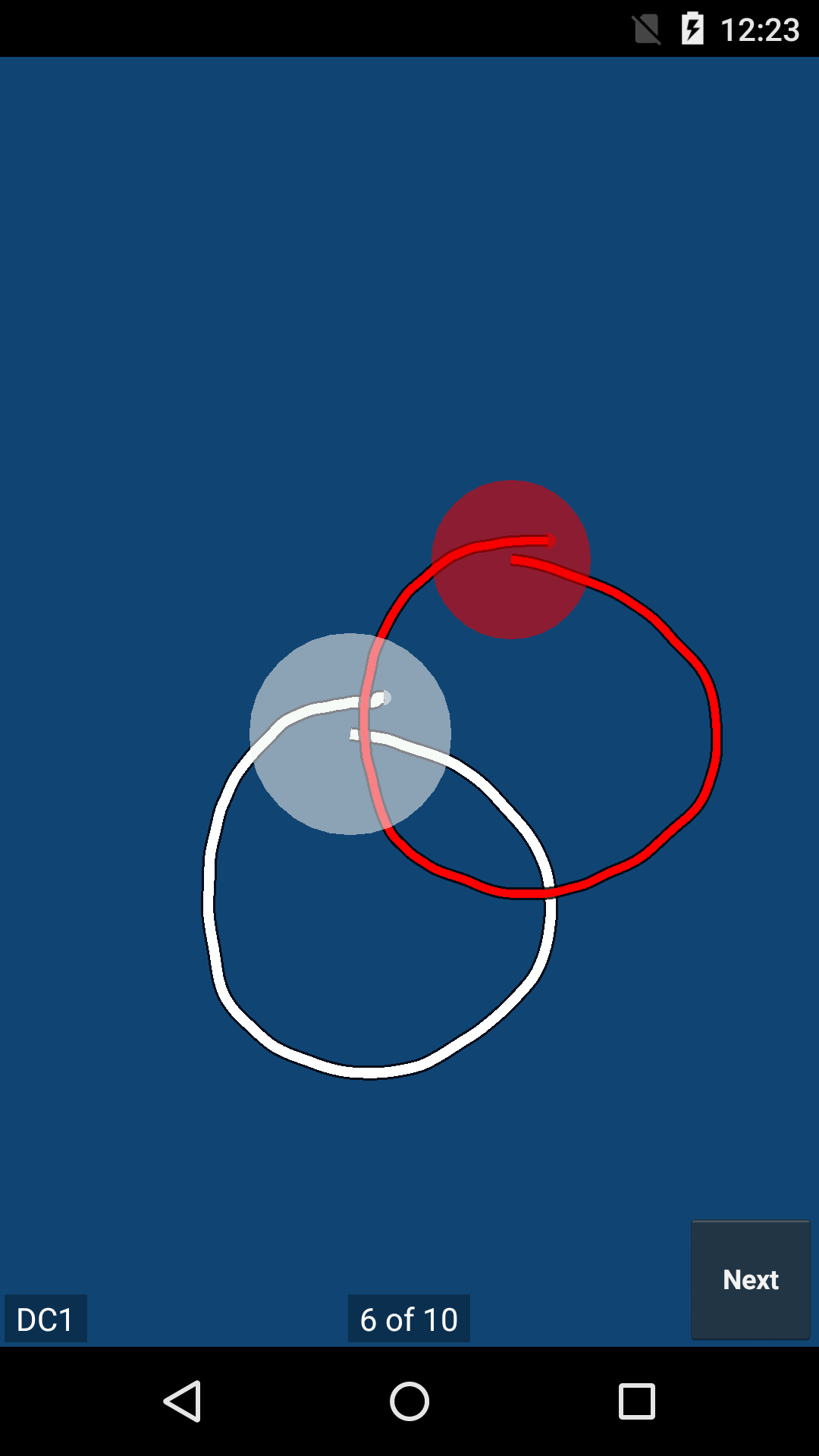}
\label{fig:inputgesturedc}
}
\hfill
\subfigure[User authentication was successful.]{
\includegraphics[width=0.29\textwidth]{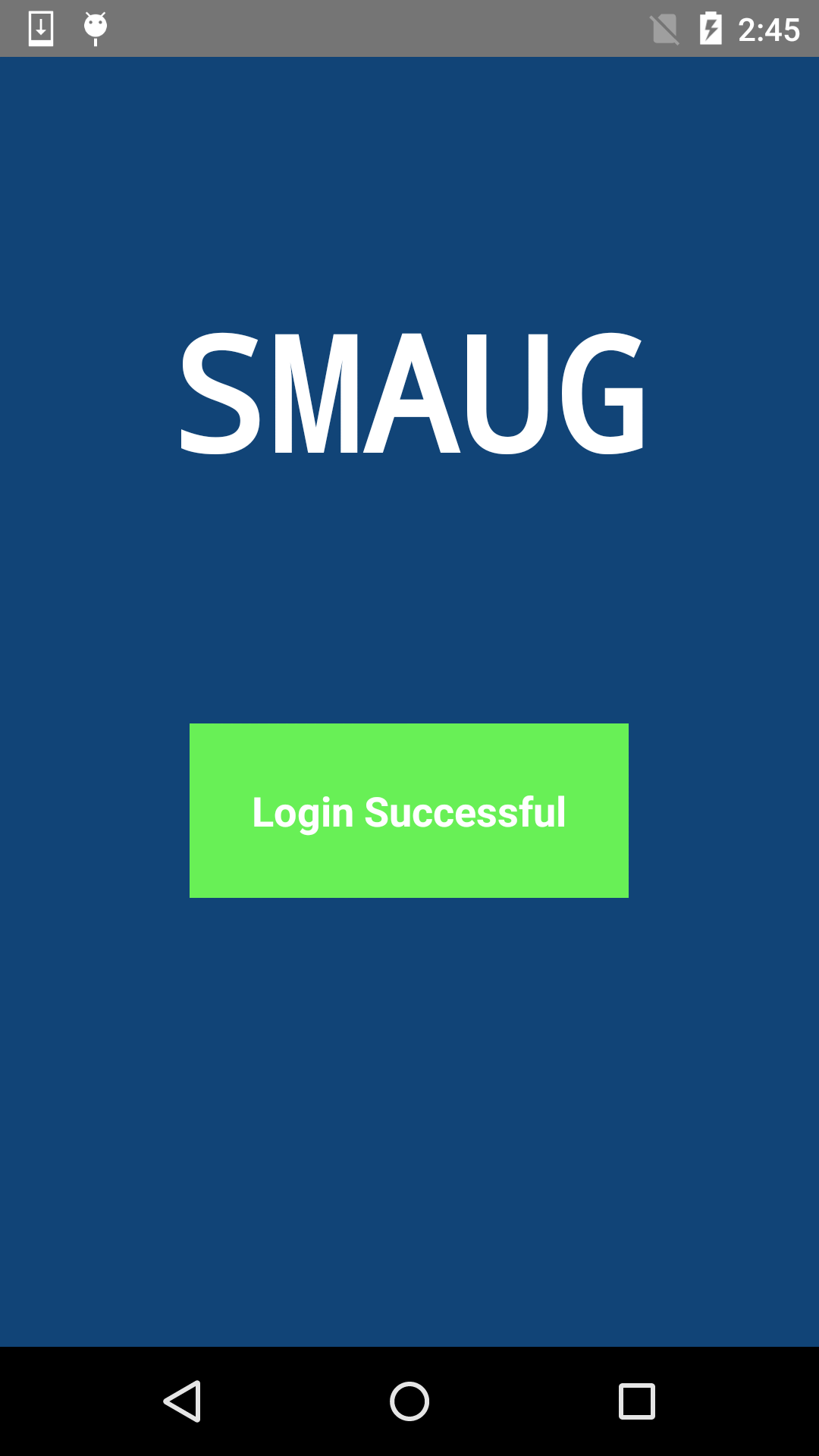}
\label{fig:loginsuccessful}
}
\vspace*{-4mm}
\caption{Input of gestures \gesture{A} and \gesture{DC} (double circle) during the enrollment phase and info message after successful authentication.}
\label{fig:inputlogin}
\end{center}
\end{figure}

For our tests, we acquired 23 test participants, ten female and thirteen male persons, for 36 test sessions.\footnote{Each participant agreed on using his/her anonymized data in this project. Furthermore our IRB staff is informed about this project.}
Out of these, 18 participants had a background in IT-security.
The participants were divided into four groups.
The first group used gestures without a background image while the second one was using gestures with a background image.
The third and fourth group had the same task, however they also performed delayed inputs, waiting from half an our up to a week between the enrollment and verification phase.
A background image helps remembering and attacking the gesture, see Section~\ref{sec:modes} and Figure~\ref{fig:inputgesturea}.
For the background images, they had the choice between a grid, a mountain view, a field of flowers, and a popular band logo.
Otherwise the choice between a uniformly black and a blue background color was given.
Each participant got a handout describing step by step how the app works and how the tests should be done.
They were also guided in the first steps by us in order to assist with the system or to answer their questions.
Usually a participant started by entering some test gestures aiming to understand the reaction of the system and \SMAUG.
Each test involved both predetermined gestures and freely chosen gestures.
With respect to the predefined gestures, they had to enter the gestures described in Table~\ref{tbl:gestures}.
\begin{table}[ht]
\centering
\begin{tabular}{l|p{90mm}}
\toprule
\textbf{Type} & \textbf{Gesture Name \& Description} \\
\midrule
\multirow{7}{*}{Single-touch} & Upper case letters \gesture{A} (see Figure~\ref{fig:inputgesturea}), \gesture{B}, \gesture{I}, \gesture{U}, and \gesture{V} \\
& the word \gesture{OK} \\
& \gesture{square} (start in corner top left, draw clockwise)\\
& \gesture{tap-rhythm} \\
& \gesture{L-Turn} (write L, turn phone left by 90 degrees, write L, turn phone right by 90 degrees) \\
\midrule
\multirow{6}{*}{Multi-touch} & \gesture{Bar} \& stripe (drawing a second line while drawing another one) \\
& \gesture{DC} (double circle; drawing two circles at the same time), see Figure~\ref{fig:inputgesturedc} \\
& \gesture{wave} (three fingers, put one after another on the touch screen and lift them in the same order) \\
\bottomrule
\end{tabular}
\vspace*{2mm}
\caption{Predefined gestures used in the experiments.}
\label{tbl:gestures}
\end{table}
For the free-form part, the participants were asked to enter a minimum of two own gestures.
Mostly letters have been entered, but also numbers, arrows, swirly lines, geometric shapes, greek letters, and also 3D-gestures (moving the device around while performing the gesture).
This once more shows the diversity in gesture generation.

For the enrollment phase, a user entered the name of the gesture and after an information screen, she started to input the first round of the gesture.
When the user has finished performing the gesture for the first round, she taps on the bottom right button ``Next'' in order to proceed to the next round.
The name of the gesture is displayed on the bottom left and the current and overall number of rounds is shown in the bottom center, see Figure~\ref{fig:inputgesturedc}.
After all ten rounds, the data gathering procedure is finished and the user is back in the menu of all stored gestures on this device.
The verification phase is structured similar, except that the user can input the gesture as often as he wants, each time yielding either a successful or failed authentication, respectively.

The gestures \gesture{A}, \gesture{I}, \gesture{U}, \gesture{V}, \gesture{square}, \gesture{wave}, \gesture{double circle}, and \gesture{L-Turn}, as well as at least one free-form gesture were entered from the participants while sitting, holding the device relaxed in portrait mode with one hand and entering with the other hand.
The gestures \gesture{A}, \gesture{B}, \gesture{OK}, \gesture{rhythm}, \gesture{bar}, and at least one free-form gesture were entered while standing relaxed while performing the inputs.

The participants also performed attacks on each other.
Recall that we allow an adversary to watch the user closely.
The user explained how he entered a gesture and immediately afterwards the attacker tried to impost the user.
For these attacks, we selected representative gestures.
Letter \gesture{I} as the most simple one consisting of one stroke and no curves, letter \gesture{A} as a one fairly easy to remember but consisting of one to three strokes, \gesture{L-Turn} as a complicated one, and \gesture{Bar} \& stripe as a simple multi-touch gesture.
Each user performed attacks on these gestures of another user.
The attacker always used the same body position as the user, i.e., sitting or standing.

\subsection{Results}\label{sec:results}
Overall, we gathered 8180 data sets, each containing the meta, touch, and motion sensor data.
For successful user authentication (true positives), we counted the number of login attempts which were necessary for the legitimate user to be successful.
77\% of all valid login attempts, including free-form gestures, were recognized by \SMAUG\ at the first try, 92\% after the second try, 99\% after the third try. Allowing a fourth login try, 100\% of all login attempts were successful.
Since this includes the free-form gestures, each participant was able to login with their self-created gesture.
As expected, the success rate also depended on how involved the gesture is. 
For example, using multi-touch gestures, the 100\% rate was already reached after two login attempts, see also Table~\ref{tbl:tvpositive} for an overview for representative gestures.
In total we registered 1490 valid login attempts.

Next, we evaluated the attack scenarios (false positives) with respect to the strong adversary model explained in Section~\ref{sec:adversarymodel}.
An impostor was detected in overall 97\% of all 610 impostor attempts.
1.3\% of these impostor attempts were due to irregular inputs of the gestures \gesture{A} and \gesture{I}.
In the 3\% of successful impostor attacks, only a single one was reproduceable once, but failed to do so over a longer time of some minutes.

In most experiments, inspecting the measured data revealed that the adversary could mimic some aspects of the gesture quite successfully but then failed miserably on other features.
These results are a strong indication for capabilities of our algorithm.
We have to stress though that this also depends on the level of complexity of the used gesture.
Most simple gestures like a single line are easier to reproduce, as mentioned for gesture \gesture{I}.
Fortunately, our experiments showed that gestures which are a little bit more involved are already sufficient for successful authentication.
For example using the gesture \gesture{A} was already sufficient to achieve a security level of 99.3\% (being the probability the gesture \gesture{A} was not attacked successfully within three attempts).

Table~\ref{tbl:tvpositive} shows the results of authentication (True Positives Rate, TPR) and attack (False Positives Rate, FPR) tests for four different, exemplary gestures.
The more complicated and unusual a gesture was, the more difficult it gets to authenticate with this gesture.
However, this holds also for the attacks which will fail with a very high probability as given in the table.
Finally, the experiments showed that the attacks failed completely for the free-form gestures, while login was still possible.
\begin{table}[!ht]
\centering\small
\begin{tabular}{c|ccc|cccc}
\toprule
\multirow{2}{*}{\textbf{Gesture}} & \multicolumn{3}{c|}{\textbf{TPR in \%}} & \multicolumn{4}{c}{\textbf{FPR in \%}} \\
 & \textbf{1st} & \textbf{2nd} & \textbf{3rd} & \textbf{1st} & \textbf{2nd} & \textbf{3rd} & \textbf{all} \\
\midrule
\gesture{I}			&77	&91&100	&7	&13 &27 &5  \\
\gesture{A}			&75	&92	&100	&0	&0	&6  &5  \\
\gesture{Bar}		&73	&93&100	&0	&0	&8	&3  \\
\gesture{FreeForm}	&65	&88	&91		&0	&0	&0	&0  \\
\midrule
all 				&77 &92 &\textbf{99} &2  &4  &8  &\textbf{3}  \\
\bottomrule
\end{tabular}
\vspace*{2mm}
\caption{TPR (True Positives Rate, successful logins) and FPR (False Positive Rate, successful attacks) for different gestures.
1st, 2nd, and 3rd, denote the number of the attempt.
Further ``all'' is the summarized result of all attempts, up to 10.
Overall after the third login attempt 98\% of users were authenticated while only 3\% of all attacks were successful.}
\label{tbl:tvpositive}
\end{table}
We also conducted ten tests with a delay of half an hour between the enrollment and the verification phase.
Remarkably, the participants who have previously performed another test got authenticated in 95\% of all login attempts using a background image, and 90\% without background image after a maximum of three login attempts.
Participants who did this test first, were authenticated in 85\% of the login tests with a background image, and 80\% without background image.
As a consequence, we see that if one gets used to a gesture as secure authentication method, the probability gets higher that the authentication will succeed in the first place.

An interesting study was to compare the same gesture from the same user in different positions, i.e., sitting and standing.
We tested this with the gesture \gesture{A}.
If the users did the enrollment phase while sitting and the authentication while standing, 89\% could login after at most three attempts, which was reproducible for most of the participants in later login attempts.
Vice versa, only 11\% were able to login while sitting, only a few managed to login after the third attempt.
When the gesture has been entered using a background image, only 10\% of the users could authenticate standing when the enrollment phase was done sitting.
Again, using a background image, 30\% of the participants could authenticate sitting when the enrollment phase was done standing.

Furthermore, we tested the viability of each combination of motion data, consisting of gyroscope, accelerometer, and fusion features.
Choosing only one data set of them did work out well for user authentication, that is in 79--86\% of all test cases.
However, an impostor was badly recognized, only in 83--88\% of all test cases.
Indeed, the combination of all of them did work best, since a small variation in input data can still be ignored.

\section{Related Work}\label{sec:RelatedWork}
As mentioned in the introduction, to the best of our knowledge no work comes with the capabilities of our authentication algorithm, even when leaving the privacy part apart.
However, there are many different fields which overlap with techniques employed in \SMAUG, also many biometric authentication schemes were inspiring our design.

\noindent\emph{Signature Verification.}
%One of the first related topics has been signature verification.
The  use of DTW for successful signature verification has been considered in \cite{DTWBVSBS}.
Follow-up work in this field copes with signature input using a pen on handheld devices \cite{Martinez-Diaz2007_SignatureHandheld}.
An overview of the current state of the art of signature verification gives \cite{DBLP:journals/tsmc/ImpedovoP08}, while new approaches of SVM are combined with signature verification in \cite{DBLP:journals/tsmc/GruberGKS10}.
Good choices of features for signatures are discussed in \cite{martinez14mobileSignRobustPerf}.
In \cite{DBLP:conf/icb/Gomez-BarreroGFOP13}, the authors investigate how signatures of the same user alter over time.
Finally, the work \cite{DBLP:conf/paams/KrishFGM13} focuses on signature verification on smartphones.

\noindent\emph{Gesture Recognition.} 
%Also related is the field of gesture recognition, which gained a lot of attention after the smart phone boom.
With respect to gesture recognition for single-touch gestures, Rubine \cite{DBLP:conf/siggraph/Rubine91} is the usual reference when comparing new single-touch algorithms.
Another prominent example for single-touch and single-stroke gesture recognition is \textdollar1 \cite{DBLP:conf/uist/WobbrockWL07}.
The authors of \cite{DBLP:conf/graphicsinterface/AnthonyW12} present a very efficient follow-up work for single-touch and multi-stroke.
In short, the algorithm joins all strokes in all possible combinations and reduces the gesture recognition problem therefore to the case of single-touch gestures.
However, this algorithm family needs a predefined set of gestures.
%Specific gesture recognizers for smart diagram environments\todo{Was ist das?} have been analyzed in \cite{DBLP:conf/grec/BickerstaffeLMM07}.
The authors in \cite{ictdbid:2292} have developed a multi-dimension DTW for gesture recognition.
In \cite{DBLP:conf/mobicom/ShahzadLS13}, the authors present a gesture based user authentication scheme for touch screens using solely the accelerometer.
3D hand gesture recognition in the air with mobile devices and accelerometer is examined in \cite{DBLP:journals/ijisec/CasanovaABS12}.
Similar research was done for Kinect and gesture recognition in \cite{DBLP:conf/ndss/TianQXW13}, also for Wii \cite{DBLP:conf/percom/LiuWZWV09}.

\noindent\emph{Sketch and Image Recognition.}
Sketches are drawings of simple symbols and recognition is similar to gesture recognition.
In \cite{DBLP:journals/cg/FieldGPRSA10}, the authors classify different sketches for technical drawings.
The work \cite{DBLP:journals/cg/KaraS05} presents a sketch recognizer for a defined set of sketches which are invariant for rotation.
An algorithm for recognition of multi-touch sketches is given in \cite{DBLP:conf/hci/SchmidtW13}.
Also, \cite{DBLP:journals/corr/NguyenYC14} shows how image recognition works and how it can be fooled.

\noindent\emph{Continuous Authentication.}
Continuous authentication means that the device constantly tracks and evaluates the inputs and movements of the user onto the device to authenticate the user.
They generally suffer fom privacy loss in some kind.
Algorithms can be found in \cite{DBLP:conf/ndss/LiZX13,DBLP:journals/tifs/FrankBMMS13,DBLP:conf/iwcmc/RybnicekLH14}.%DBLP:conf/trustcom/FengZCS13

\noindent\emph{Mobile Authentication and Graphical Passwords.} 
In \cite{DBLP:conf/css/GaoJLL13}, the authors present an attack on the graphical password system of Windows 8.
\cite{DBLP:journals/jsw/GaoJYM13} gives an overview of graphical password schemes developed so far.
An enhancement for the Android pattern authentication was presented in \cite{DBLP:conf/chi/LucaHBLH12}, which utilizes the accelerometer.
The authors of \cite{DBLP:conf/cns/Sun0ZZ14} give an authentication algorithm where up to five fingers can be used for multi-touch single-stroke (per finger) in combination with touch screen and accelerometer.
Furthermore, they defined the adversary models for mobile gesture recognition based on \cite{DBLP:conf/ndss/TianQXW13}, which are all weaker than our adversary model.
In \cite{DBLP:conf/mobisys/ShermanCYSMLOR14} the authors allow multi-touch and free-form gestures to measure the amount of information of a gesture which can be used for authentication.
Finally, \cite{DBLP:conf/chi/Sae-BaeAIM12} presents a multi-touch authentication algorithm for five fingers using touch screen data and a predefined gesture set.
In \cite{DBLP:conf/chi/YangCLO16} the authors test free-form gesture authentication in non-labor environments.

\section{Conclusion}\label{sec:Conclusion}
We presented \SMAUG, a novel authentication scheme for mobile devices based using gestures.
The gestures can be freely chosen, may consist of multiple strokes of different length and can be drawn using multiple fingers at the same time.
Moreover, the layout of the gesture is freely chosen by the user and is in no way restricted by \SMAUG.
To tie these gestures to a user, not only the form of the gesture itself is considered but also how it has been entered.
Hence, we combine \emph{knowledge} of the form of the gesture as well as the \emph{biometric} properties of its input resulting in a two-factor authentication.
Data from different sensors is collected to identify individual biometric properties of the user.
We stress that only sensors are used that are commonly found in mobile devices: touch screen, gyroscope, and accelerometer. 
Compared to existing work, our scheme provides a significant higher level of flexibility and achieves security against the strongest possible attacker -- being non-continuous and without any loss of privacy at the same time.
In this work, we also provide an implementation which allowed us to perform experiments with various participants.
Our experiments show that while legitimate users are correctly authenticated with overwhelming probability of 99\% after the third attempt, an attacker fails to impersonate the user with overwhelming probability of 97\%, even if he knows what gesture is required and how it has to be entered. 
Hence, we believe that \SMAUG\ provides an important contribution towards solving the long-standing open problem of providing usable and secure authentication for mobile devices.

Although our scheme achieves already very good and reliable results, we see several possible directions for further improvements.
Our experiments were carried out in labor environments, i.e., the user had a fixed pose during enrollment and verification.
In practice, one would require more robust schemes which on the one hand also work while the user is moving, i.e., walking or driving, and on the other hand need to respect rotation of the device, i.e., lying on a table or when the user is lying.
However, implementing this will reduce security.
Currently \SMAUG\ is strictly forcing the user to enter the strokes always in the same order.
While this makes sense in general, in some cases, e.g., when using a pinch gesture, this may be hard to achieve.
Here, a small time window which allows for changes of the stroke order may be helpful.
The user has to repeat the gesture ten times during enrollment phase.
One may argue that enrollment should only rarely take place, but it would nonetheless be advantageous to reduce this number.
Finally, as mentioned in the related work, biometric input may alter over time due to the user's natural behavior.
Combining this with \SMAUG\ is an open problem which we plan to tackle in the future.
%Future Work
%- optimization space and computation time
%- Smartphone is lying on the table or is rotated during input
%- user is moving (walking, driving, ...)
%- Pointer identifier and pointer number may vary
%- reduce number of rounds during enrollment phase
%- implementation on Nexus instead of computer
%- design choices experimental -> are there better choices?

\bibliographystyle{abbrv}
\bibliography{references}

\newpage
\begin{sidewaystable}[p]
\renewcommand{\arraystretch}{1.2}
\centering
\begin{tabular}{c|ccccccccccc}
\toprule
 & \textbf{Multi-} & \textbf{Free-Form} & \textbf{Multi-} & \textbf{Multi-} & \textbf{Multi-} & \textbf{Multi-} & \textbf{Sensor} & \textbf{Graphical} & \textbf{Continuous} & \textbf{Operating} & \textbf{Sample} \\
\textbf{Work} & \textbf{Gesture} & \textbf{Gesture} & \textbf{Touch} & \textbf{Stroke} & \textbf{Factor} & \textbf{Sensor} & \textbf{Fusion} & \textbf{Password} & \textbf{Authentication} & \textbf{System} & \textbf{Size} \\
\midrule
\cite{DTWBVSBS} & \ctableno & \ctableyes & \ctableno & \ctableyes & \ctableno & \ctableno\ (T) & \ctableno & \ctableno & \ctableyesS no & \ctablemaybe & \ctablemaybeS 20 \\
\cite{DBLP:conf/nordichi/WeissL08} & \ctableno & \ctableno & \ctableno & \ctableno & \ctableno & \ctableno\ (T) & \ctableno & \ctableno & \ctableyesS no & \ctableyesS any & \ctablemaybeS 24 \\
\cite{DBLP:conf/percom/LiuWZWV09} & \ctableno & \ctableyes & \ctableno & \ctableno & \ctableno & \ctableno\ (A) & \ctableno & \ctableno & \ctableyesS no & \ctablemaybeS Wii & \ctablenoS ? \\
\cite{DBLP:conf/btas/Sae-BaeMI12} & \ctableno & \ctableno & \ctablemaybeS yes (5fix) & \ctableno & \ctableno & \ctableno\ (T) & \ctableno & \ctableno & \ctableyesS no & \ctableyesS any & \ctableyesS 10 \\
\cite{DBLP:conf/chi/Sae-BaeAIM12} & \ctableno & \ctableyes & \ctablemaybeS yes (5fix) & \ctableno & \ctableno & \ctableno\ (T) & \ctableno & \ctableno & \ctableyesS no & \ctableyesS any & \ctableyesS 10 \\
\cite{DBLP:conf/chi/LucaHBLH12} & \ctableno & \ctableno & \ctablemaybeS partly (2) & \ctableno & \ctableyes & \ctablemaybeS yes (TA) & \ctableno & \ctableno & \ctableyesS no & \ctablemaybeS Android & \ctablemaybeS 20 \\
\cite{DBLP:conf/iccnc/ZhuWWZ13} & \ctablemaybeS partly & \ctableno & \ctableno & \ctableyes & \ctableno & \ctablemaybeS yes (AGM) & \ctableno & \ctableno & \ctablenoS yes & \ctablenoS modified & \ctableyesS -- \\
\cite{DBLP:conf/mobicom/ShahzadLS13} & \ctableyes & \ctableno & \ctableyes\ (5) & \ctableno & \ctableno & \ctablemaybeS yes (TA) & \ctableno & \ctableno & \ctableyesS no & \ctableyesS any & \ctablemaybeS 25 \\
\cite{DBLP:conf/ndss/LiZX13} & \ctablemaybeS partly & \ctableno & \ctableno & \ctableyes & \ctableno & \ctableno\ (T) & \ctableno & \ctableno & \ctablenoS yes & \ctablenoS modified & \ctableyesS -- \\
\cite{DBLP:journals/tifs/Sae-BaeMIA14} & \ctableno & \ctableyes & \ctableyes\ (5) & \ctableno & \ctableno & \ctableno\ (T) & \ctableno & \ctableyes & \ctableyesS no & \ctableyesS any & \ctableyesS 10 \\
\cite{DBLP:conf/mobisys/ShermanCYSMLOR14} & \ctableno & \ctableyes & \ctableyes\ (4) & \ctableno & \ctableno & \ctableno\ (T) & \ctableno & \ctableno & \ctableyesS no & \ctableyesS any & \ctableyesS 10 \\
\cite{DBLP:conf/cns/Sun0ZZ14} & \ctableno & \ctableyes & \ctableyes\ (5) & \ctableno & \ctableno & \ctablemaybeS yes (TA) & \ctableno & \ctableno & \ctableyesS no & \ctableyesS any & \ctablemaybeS 20-50 \\
\cite{DBLP:conf/icnp/ZhengBHW14} & \ctableno & \ctableno & \ctableno & \ctableno & \ctableyes & \ctablemaybeS yes (TA) & \ctableno & \ctableno & \ctableyesS no & \ctableyesS any & \ctablemaybeS 20 \\
\midrule
\SMAUG & \ctableyes & \ctableyes & \ctableyes\ (10+) & \ctableyes & \ctableyes & \ctableyes\ (TAG) & \ctableyes & \ctableyes & \ctableyesS no & \ctableyesS any & \ctableyesS 10\\
\bottomrule
\end{tabular}
\caption{Overview and comparison of mobile authentication schemes and their features.
For multi-touch, the number in brackets denotes the possible maximum numbers of fingers used at the same time, ``fix'' means that this exact amount of fingers must be used
For multi-sensor, T denotes touch sensor, A accelerometer, G gyroscope, and M magnetometer.
A (red) ``\ctableno'' denotes that the work of this specific row lacks this feature, and vice versa a (green) ``\ctableyes'' displays that this feature is fully supported.
The colors of ``\ctableyes'' and ``\ctableno'' are interchanged for continuous authentication due to the reduced privacy.
As the table shows, our algorithm \SMAUG\ is rich on features and steps ahead of existing work, providing the richest set of capabilities.}
\label{tbl:authschemes}
\end{sidewaystable}
\end{document}